\documentstyle[preprint,aps,epsfig]{revtex}
\begin{document}
\draft

\title{Probing small-$x$ gluons 
by low-mass Drell-Yan pairs at colliders}

\author{George Fai\dag, Jianwei Qiu\ddag, and Xiaofei Zhang\dag\  }

\address{\dag\  CNR, Department of Physics,
Kent State University, Kent, Ohio 44242, USA}

\address{\ddag\ Department of Physics and Astronomy, Iowa State University,
Ames, Iowa 50011, USA}


\maketitle

\vspace*{-6.5cm}
\begin{flushright}
{KSUCNR-204-04}
\end{flushright}
\vspace*{5.5cm}

\begin{center}
\today
\end{center}

\begin{abstract}
The transverse-momentum ($Q_T$) distribution of low-mass Drell-Yan pairs 
is calculated in QCD perturbation theory with all-order resummation
of $\alpha_s (\alpha_s \ln(Q^2_T/Q^2))^n$ type terms. 
We demonstrate that the rapidity distribution of low-mass Drell-Yan
pairs at large-enough transverse momentum is an advantageous source of
constraints on the gluon distribution and its nuclear dependence.  
We argue that low-mass Drell-Yan pairs in the forward region provide 
a good and clean probe of small-$x$ gluons at collider energies.
\end{abstract}

\vskip .5 cm

\pacs{12.38.Cy, 25.75.Dw}


\section{Introduction}
\label{sec:intro}

Precise knowledge of the gluon distribution at small $x$ is critical 
for reliable predictions of hard processes in hadron-hadron,
hadron-nucleus, and nucleus-nucleus collisions at collider energies.
Although tremendous effort has been invested in the study of the gluon
distribution of a free hadron, there are still large uncertainties in
the distribution in both the large and small $x$ regions 
\cite{Huston:1998jj,Pumplin:2002vw,Martin:2003sk,Eskola:2002yc}.
Because of the limited kinematic range that can be probed by fixed target
arrangements, we have not had enough data to shed light on the gluon
distribution in a nucleus. Thus, information on the gluon distribution 
in a nucleus is far from complete 
\cite{Qiu:1986wh,Frankfurt:xz,Eskola:1992zb,Eskola:1998df,Eskola:1998iy,Hirai:2001np,Huang:1997ii,Li:2001xa,Frankfurt:2002kd,Eskola:2002us,Armesto:2002ny,Accardi:2003be}.  
With better data from collider facilities, like
the Relativistic Heavy Ion Collider (RHIC) and
the future Large Hadron Collider (LHC), and also from
the Tevatron at Fermilab,
we expect much better 
information on the gluon distribution in free hadrons as well as in nuclei.
In this paper, we demonstrate that
the rapidity distribution of low-mass Drell-Yan pairs at large-enough
transverse momentum is an advantageous source of constraints on the
gluon distribution and its nuclear dependence.  

A good process from the point of view of probing the gluon distribution 
must satisfy the following criteria: (i) it has to be reliably calculable 
within the factorization frame work of perturbative Quantum Chromodynamics
(pQCD); (ii) its production cross section must be dominated by
gluon-initiated sub-processes; and (iii) it should have sufficient
production rate for the process to be observed.  For probing the gluon 
distribution in a nucleus, we need to add another criterion: (iv)
the process should have relatively small initial- and final-state medium 
effects because of the twist-2 nature of the gluon distribution.  

It was first pointed out by Berger et al. \cite{Berger:1998ev} that 
the transverse-momentum distribution of Drell-Yan (massive) lepton
pairs produced in hadronic collision is an advantageous source of
constraints on the gluon distribution because of the dominance of gluon
initiated subprocesses. These authors also emphasized that this
process is free from the complications of photon isolation that beset
studies of prompt photon production\cite{Berger:1998ev,Berger:1990es}.   
Due to the ``Drell-Yan factor'' in the cross section (the first
factor on the right-hand side of Eq. (\ref{DY-Vph}) below), it is 
useful to focus on low-mass Drell-Yan pairs to satisfy condition (iii). 
This, however, necessitates a recently-introduced resummation 
procedure\cite{Berger:2001wr}.
 
In this paper we study the rapidity and transverse-momentum distribution
of low-mass Drell-Yan pairs at RHIC, 
Tevatron
and LHC energies, and 
the nuclear dependence of these distributions. (A short account of 
some aspects of the present work was given earlier\cite{Fai:2004qu}.)  
In Sec.~\ref{sec:frame}
the theoretical framework of our calculation is briefly reviewed.
In Sec.~\ref{sec:glu}, we show that if the transverse momentum 
$Q_T$ is larger than the invariant mass $Q$ of the pair,
the gluon 
initiated
subprocesses give more
than 80\% of the low-mass Drell-Yan cross section at these colliders.
We present our calculations for the cross sections of massive
lepton-pair production at these collider energies in Sec.~\ref{sec:X}, 
and demonstrate that low-mass Drell-Yan lepton pairs are measurable. 
We also show that the gluon dominance does not reduce for all relevant 
rapidities. In Sec.~\ref{sec:xrange}, we study the range of parton momentum 
fraction $x$ that can be probed at these colliders while we vary the rapidities.  
We find that the forward region in rapidity is an excellent place to probe
small-$x$ gluons. In Sec.~\ref{sec:nucl}, we calculate the  nuclear modification
ratio, $R_{AB}$, for $dAu$, $AuAu$, $pPb$ and $PbPb$ collisions.  We investigate the 
effect of isospin for heavy ion beams, and the effect of nuclear
modifications on parton distribution functions.  We find that isospin plays an 
important role and study its impact on the $R_{AB}$ ratio. We also
observe an interesting dependence and constraint on the effective nuclear
parton distributions. We point out that the forward suppression of Drell-Yan
pair production in nucleus-nucleus collisions at LHC energies is expected 
to show a very different behavior from the pattern at RHIC. Our conclusions 
are given in Sec.~\ref{sec:concl}.

\section{Calculational framework}
\label{sec:frame}

The massive lepton pair of the Drell-Yan process is produced via the 
decay of an intermediate virtual photon, $\gamma^*$. Within the
context of pQCD, the Drell-Yan cross section in a collision between
hadrons $A$ and $B$, 
$A(P_A)+B(P_B)\rightarrow \gamma^*(\rightarrow l\bar{l}(Q))+X$,
can be expressed in terms of the cross section for production of an 
unpolarized virtual photon of the same invariant mass\cite{Berger:1998ev}, 
\begin{equation}
\frac{d\sigma_{AB\rightarrow \ell^+\ell^-(Q) X}}{dQ^2\,dQ_T^2\,dy}
= \left(\frac{\alpha_{em}}{3\pi Q^2}\right)
  \frac{d\sigma_{AB\rightarrow \gamma^*(Q) X}}{dQ_T^2\,dy}\, \,\, ,
\label{DY-Vph}
\end{equation}
with 
\begin{eqnarray}
\frac{d\sigma_{AB\rightarrow \gamma^*(Q) X}}{dQ_T^2\,dy}
&=&\sum_{a,b} \int dx_1 \phi_{a/A}(x_1,\mu) 
              \int dx_2 \phi_{b/B}(x_2,\mu)
\nonumber\\
&\ & \times 
     \frac{d\hat{\sigma}_{ab\rightarrow \gamma^*(Q) X}}
          {dQ_T^2\,dy}
          (x_1,x_2,Q,Q_T,y;\mu)\, \,\, ,
\label{DY-fac}
\end{eqnarray}
where $\Sigma_{a,b}$ runs over all parton flavors; the variables $Q$,
$Q_T$, and $y$ are the invariant mass, transverse momentum, and
rapidity of the pair, respectively; $X$ stands for an
inclusive sum over final states that recoil against the virtual
photon. In Eq.~(\ref{DY-fac}), 
the $\phi_{a/A}$ and $\phi_{b/B}$ are parton distribution
functions, $\mu$ represents both the renormalization and the factorization 
scales (taken to be equal), and the 
$d\hat{\sigma}_{ab\rightarrow \gamma^*(Q) X}/dQ_T^2\,dy$
are short-distance hard parts calculated perturbatively order-by-order
in powers of $\alpha_s$.  In Eq.~(\ref{DY-Vph}), an integration has
been performed over the angular distribution in the lepton-pair rest
frame. Because of the Drell-Yan factor, 
$\alpha_{em}/(3\pi Q^2)$ in Eq.~(\ref{DY-Vph}), the Drell-Yan cross 
section suffers from a low production rate\cite{Berger:2001wr}. 

To increase the production rate, it is desired to measure low-mass
lepton pairs.  On the other hand, the perturbatively calculated
partonic hard parts,
$d\hat{\sigma}_{ab\rightarrow \gamma^*(Q) X}/dQ_T^2\,dy$ in
Eq.~(\ref{DY-fac}), have $\alpha_s (\alpha_s \ln(Q^2_T/Q^2))^n$ type
large logarithms if $Q^2\ll Q^2_T$, and the process has a potential of 
large background from open charm (or bottom) decay.  Recently, it was
shown\cite{Berger:2001wr} that the perturbatively factorized Drell-Yan 
cross section in Eq.~(\ref{DY-fac}) can be systematically re-organized 
into a new factorized form so that the  
$\alpha_s (\alpha_s \ln(Q^2_T/Q^2))^n$ type large logarithms are
resummed to all orders in $\alpha_s$:
\begin{eqnarray}
\frac{d\sigma_{AB\rightarrow \gamma^*(Q) X}}{dQ_T^2\,dy}
&=&\sum_{a,b} \int dx_1 \phi_{a/A}(x_1,\mu) 
              \int dx_2 \phi_{b/B}(x_2,\mu)
\nonumber\\&\ &
 \times \Bigg\{ 
 \frac{d\hat{\sigma}_{ab\rightarrow \gamma^*(Q) X}^{(\rm Dir)}}
      {dQ_T^2\,dy}
      (x_1,x_2,Q,Q_T,y;\mu_{\rm Fr},\mu)
\label{DY-mfac} \\
&\ & {\hskip 0.1in} +
\sum_c \int\frac{dz}{z^2} \left[
 \frac{d\hat{\sigma}_{ab\rightarrow c X}^{(\rm F)}}{dQ_{T_c}^2\,dy}
        (x_1,x_2,Q_c=\frac{\hat{Q}}{z};\mu_{\rm Fr},\mu) \right]\nonumber\\
&\ & {\hskip 0.2in} \times
D_{c\rightarrow\gamma^*(Q) X}(z,\mu_{\rm Fr}^2;Q^2)
 \Bigg\} \, \,\, ,
\nonumber 
\end{eqnarray}
where $\Sigma_{a,b}$ and $\Sigma_c$ run over all parton flavors.
In Eq.~(\ref{DY-mfac}), 
$d\hat{\sigma}_{ab\rightarrow \gamma^*(Q) X}^{(\rm Dir)}/dQ_T^2\,dy$
and $d\hat{\sigma}_{ab\rightarrow c X}^{(\rm F)}/dQ_{T_c}^2\,dy$, 
are re-organized perturbatively calculated short-distance hard parts
with the superscripts (Dir) and (F) indicating the direct and
fragmentation contributions, respectively; 
the $D_{c\rightarrow\gamma^*(Q) X}$ are virtual photon
fragmentation functions \cite{Qiu:2001nr}, which include all order
resummation of the $\alpha_s (\alpha_s \ln(Q^2_T/Q^2))^n$ type large
logarithms. The quantity $\mu_{\rm Fr}$ represents the fragmentation scale.  
The four-vector $\hat{Q}^\mu$ is defined to be $Q^\mu$ but with $Q^2$
set to be zero. 
The CTEQ5M parton distribution 
functions (PDFs) are used in our calculations reported here, 
with all scales set to $\sqrt{Q_T^2+Q^2}$. This combination of the 
physical scales in the problem appears very natural and has 
appropriate limits when $Q^2_T \ll Q^2$ and when  $Q^2\ll Q^2_T$.
We are primarily interested in the latter case in the present work. 

As pointed out in Ref.~\cite{Berger:2001wr}, 
the resummation of the $\alpha_s (\alpha_s \ln(Q^2_T/Q^2))^n$ type large
logarithms is to re-organize the perturbative  
short-distance hard part, 
$d\hat{\sigma}_{ab\rightarrow \gamma^*(Q) X}/dQ_T^2\,dy$
in Eq.~(\ref{DY-fac}) into a ``direct'', 
$d\hat{\sigma}_{ab\rightarrow \gamma^*(Q) X}^{(\rm Dir)}/dQ_T^2\,dy$, 
and a ``fragmentation'' contribution, 
$d\hat{\sigma}_{ab\rightarrow c X}^{(\rm F)}/dp_{T_c}^2\,dy$,
(Eq.~(\ref{DY-mfac})).
It is important to realize that the direct and fragmentation contributions
each have their separate perturbative expansions, recasting the 
single expansion of the conventional QCD factorization approach into 
two series\cite{Berger:2001wr}. 
A significant advantage of this re-organization 
is that the direct and fragmentation expansions are evaluated at a single 
hard scale and are free of large logarithms, and consequently, much more
stable perturbatively.

\section{Gluon-initiated contribution 
}
\label{sec:glu}

To study the relative size of gluon-initiated contributions, we 
define the ratio
\begin{equation}
R_g = \left.
\frac{d{\sigma}_{AB\rightarrow \gamma^*(Q) X}(\mbox{gluon-initiated})}
     {dQ_T^2\,dy} \right/
\frac{d{\sigma}_{AB\rightarrow \gamma^*(Q) X}}{dQ_T^2\,dy}\, \,\, .
\label{R-g}
\end{equation}
The numerator includes the contributions from all partonic sub-processes with 
at least one initial-state gluon, and the denominator includes 
all sub-processes. 
In Fig.~1(a), we show $R_g$ as a function of $Q_T$ at fixed $y$, while
Fig.~1(b) displays $R_g$ as a function of $y$ at fixed $Q_T$  
at $Q=2$~GeV for proton-proton collision at 
RHIC energy.
\begin{figure}[h]
\begin{minipage}[c]{7.8cm}
\centerline{\includegraphics[height=7cm,width=8cm]{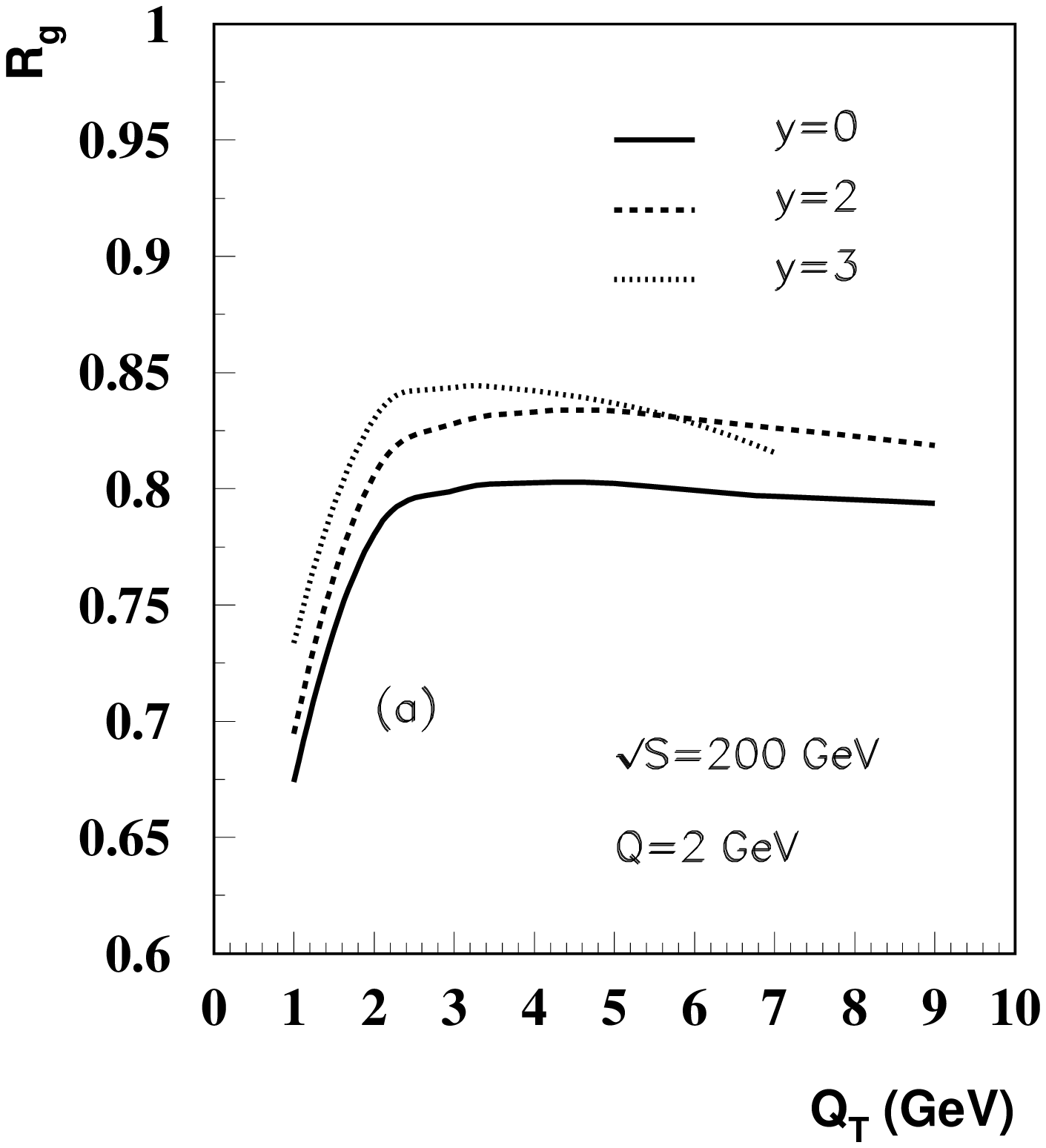}} 
\label{fig1}
\end{minipage}
\hfill
\begin{minipage}[c]{7.8cm}
\includegraphics[width=8cm,height=7cm]{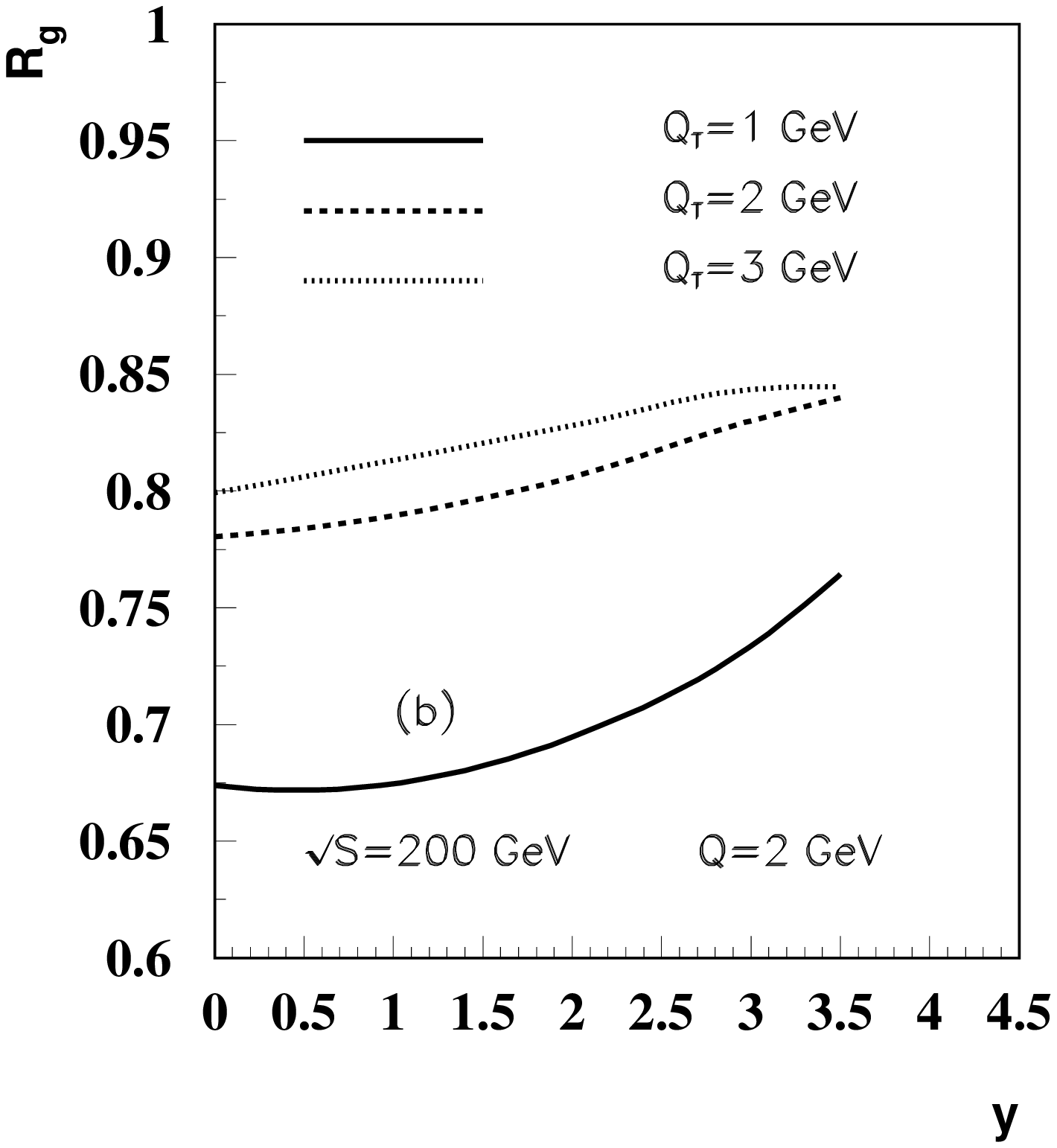} 
\vspace{-0.3in}
\end{minipage}
\hfill
\vspace{0.2in}
\caption{The ratio $R_{g}$ defined in Eq.~(\protect\ref{R-g})
with $Q=$ 2~GeV at $\sqrt{s}=200$ GeV.
Part (a): $R_{g}$ as a function of $Q_T$ for different
rapidities; part (b): $R_{g}$ as a 
function of rapidity $y$ for different transverse momenta.}
\end{figure}
\noindent Figures 2 and 3 contain similar information 
for proton-antiproton collision at Tevatron energies,
$\sqrt s=630$ GeV and $\sqrt s=1960$ GeV, respectively. 

The $Q_T$ dependence of $R_g$ shows a rapid increase up to $Q_T \approx Q$, 
followed by a plateau with a value at around (or above) 0.8 at all energies. 
This confirms that, for sufficiently large $Q_T$, gluon-initiated sub-processes 
dominate the Drell-Yan cross 
section and thus low-mass Drell-Yan lepton-pair production at large transverse momentum 
is an excellent source of information on the gluon distribution at RHIC, Tevatron and LHC 
energies\cite{Berger:1998ev}. The fall-off of $R_g$ towards very large $Q_T$ is related 
to the reduction of phase space and to the fact that the cross sections are evaluated at larger 
values of the parton momentum fractions.  
As a function of rapidity, $R_g$ increases slightly
as one moves in the forward direction (in particular at RHIC energy,
see parts (b) of the Figures).
This already
calls attention to the forward region as a good kinematical domain to test gluon 
distributions. 
An example of the energy dependence of $R_g$ is displayed in Fig.~4. 
We see that $R_g$ increases slightly as $\sqrt s$ increases, but
the dependence is not too strong. (Note the logarithmic scale for $\sqrt{s}$.)  

\begin{figure}
\begin{minipage}[c]{7.8cm}
\centerline{\includegraphics[height=7cm,width=8cm]{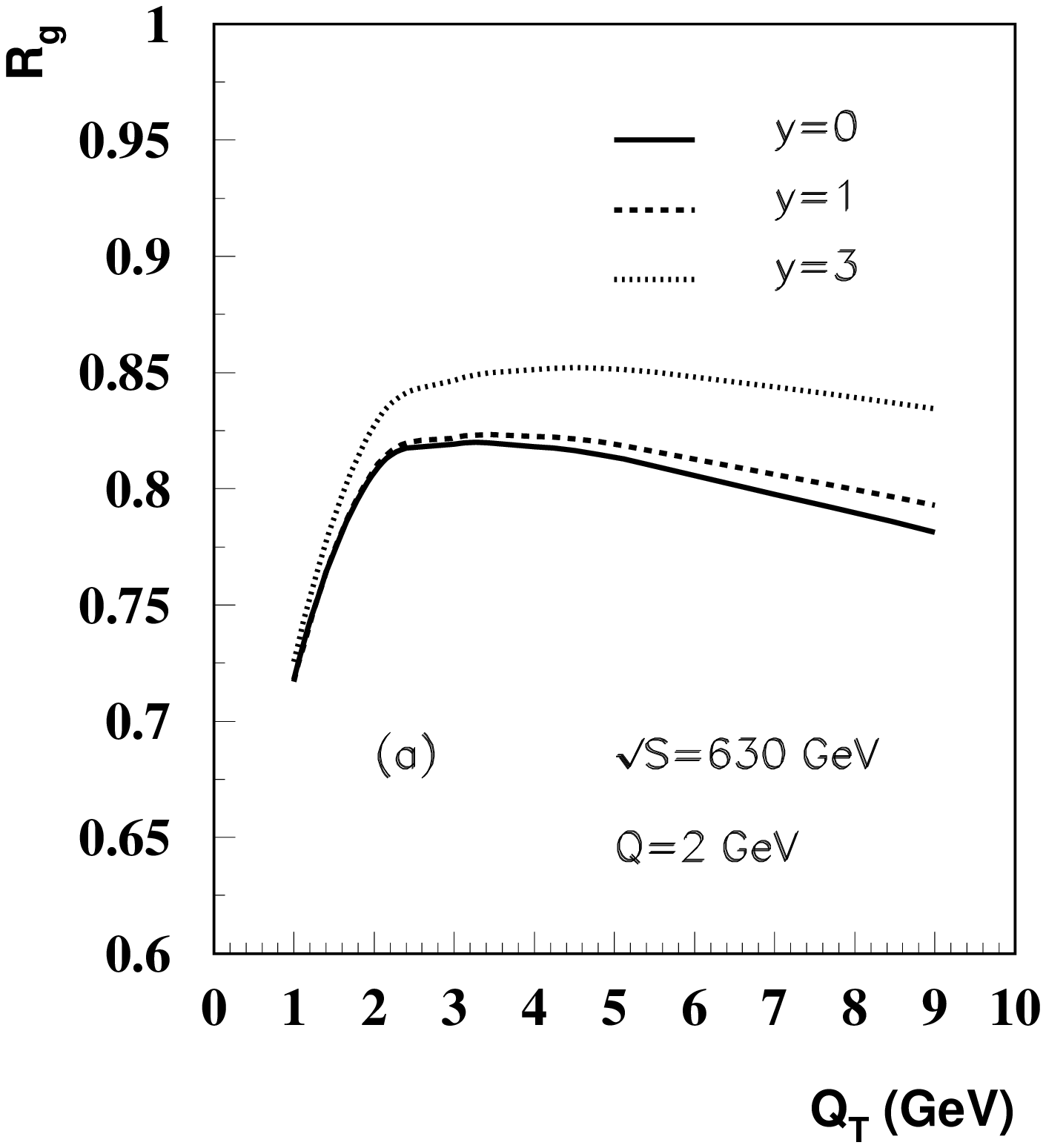}} 
\label{fig2}
\end{minipage}
\hfill
\begin{minipage}[c]{7.8cm}
\includegraphics[width=8cm,height=7cm]{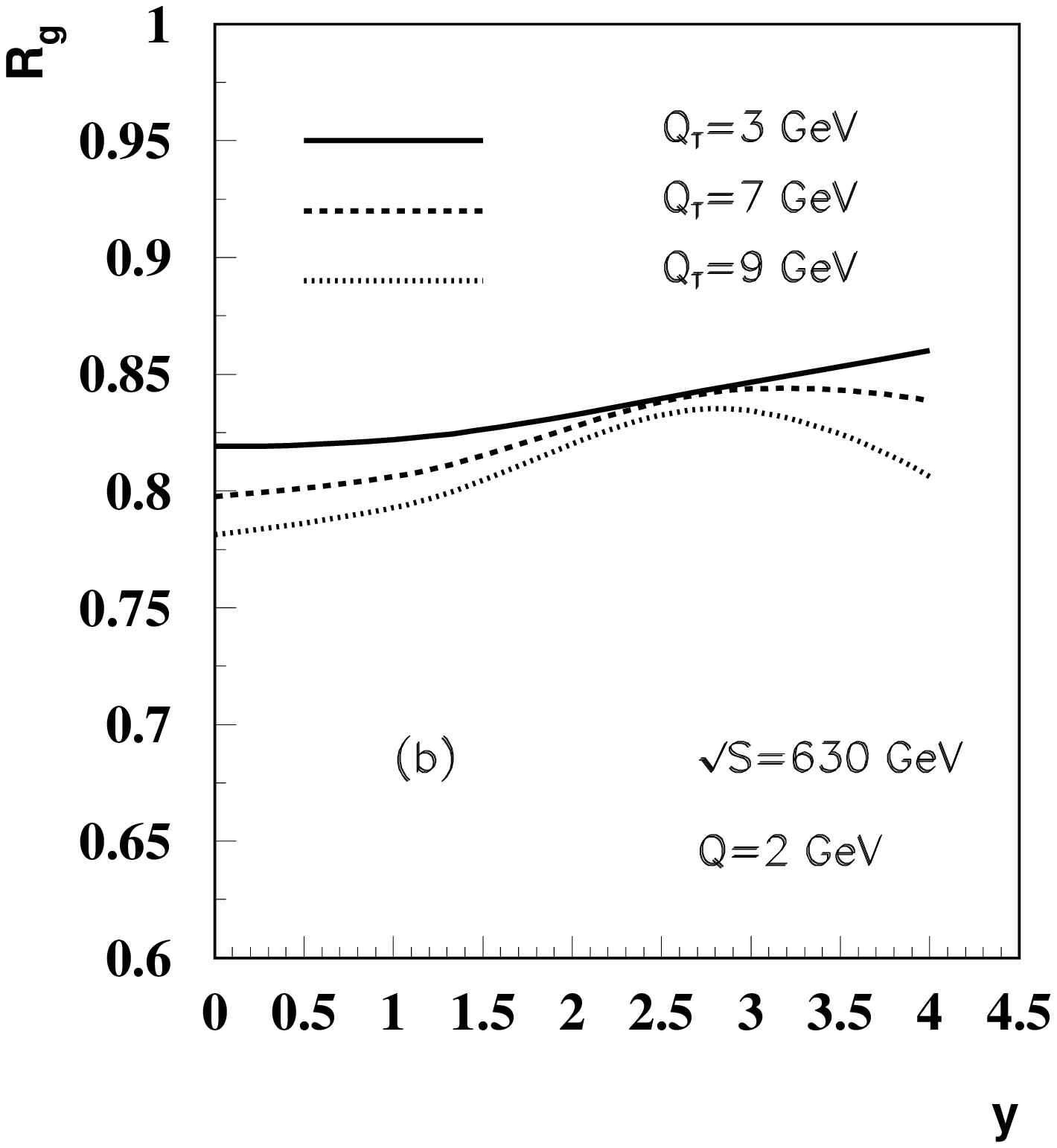} 
\vspace{-0.3in}
\end{minipage}
\vspace{0.2in}
\caption{The ratio $R_{g}$ defined in Eq.~(\protect\ref{R-g})
with $Q=$ 2~GeV at $\sqrt{s}=630$ GeV. Part (a): $R_{g}$ as a function 
of $Q_T$ for different rapidities; part (b): $R_{g}$ as a 
function of rapidity $y$ for different transverse momenta.}
\end{figure}

\begin{figure}
\vspace{0.4in}
\begin{minipage}[c]{7.8cm}
\centerline{\includegraphics[height=7cm,width=8cm]{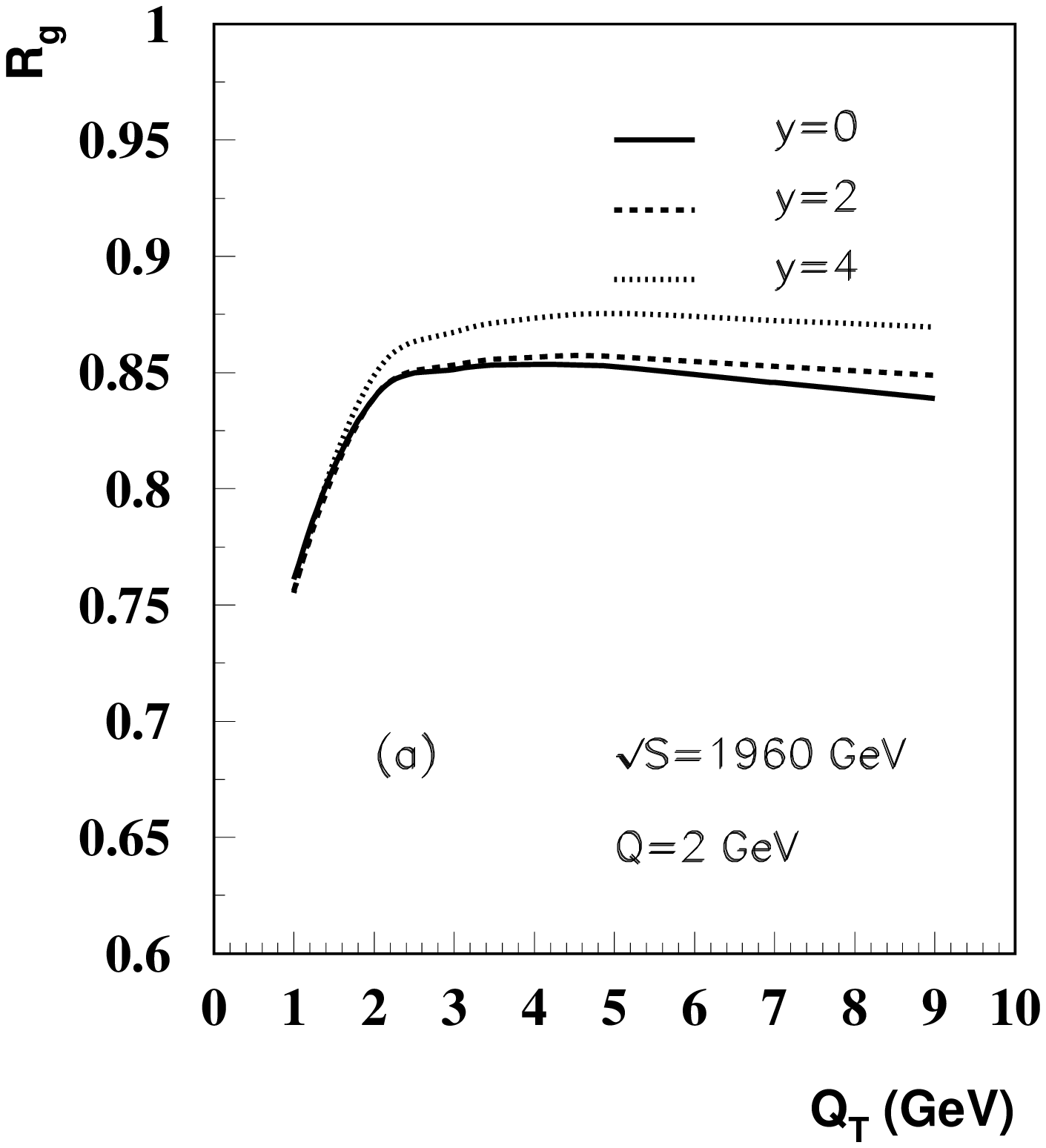}} 
\label{fig3}
\end{minipage}
\hfill
\begin{minipage}[c]{7.8cm}
\includegraphics[width=8cm,height=7cm]{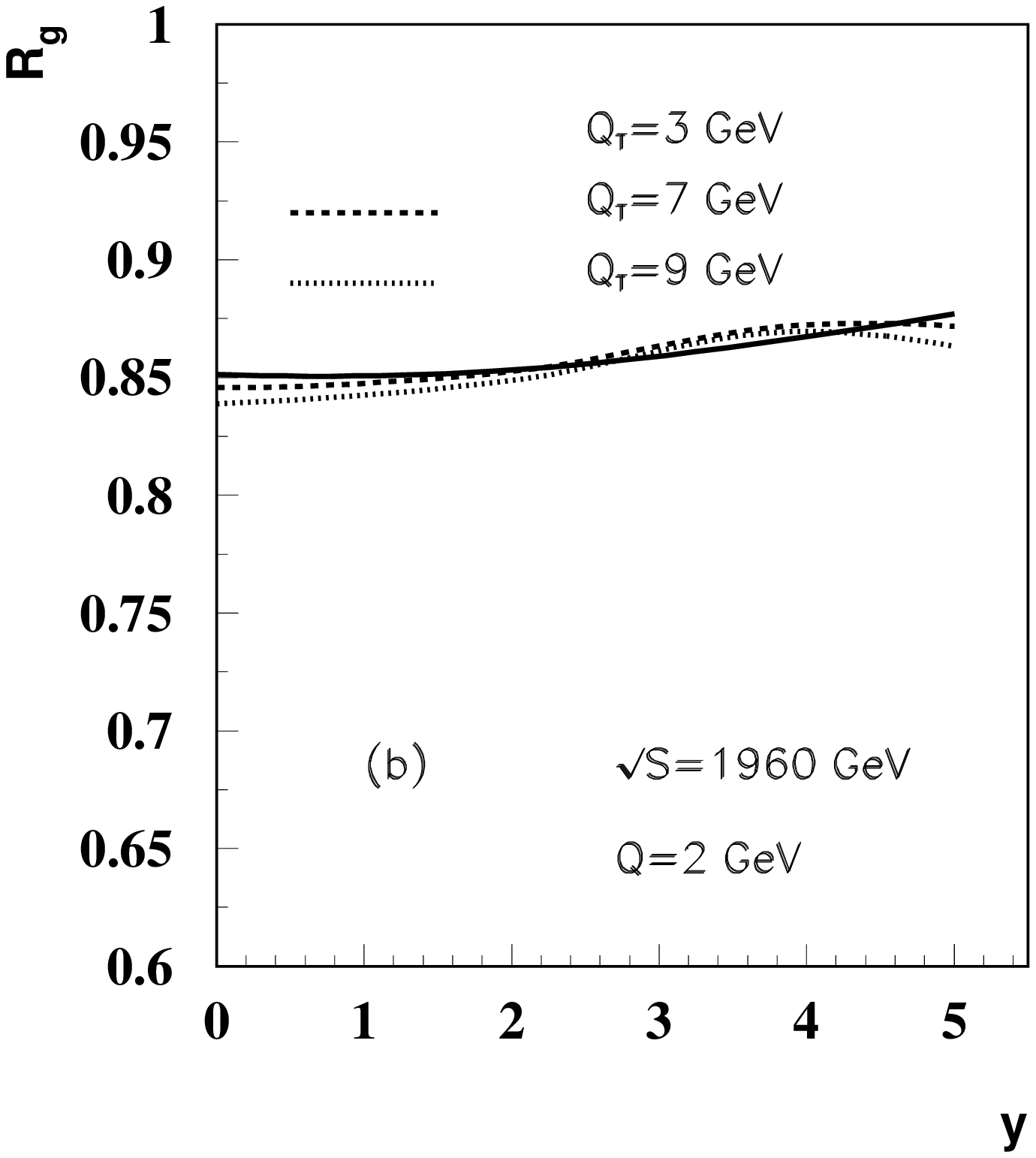} 
\vspace{-0.3in}
\end{minipage}
\vspace{0.2in}
\caption{The ratio $R_{g}$ defined in Eq.~(\protect\ref{R-g})
with $Q=$ 2~GeV at $\sqrt{s}=1.96$ TeV. Part (a): $R_{g}$ as a
function of $Q_T$ for different rapidities; part (b): $R_{g}$ as a 
function of rapidity $y$ for different transverse momenta.}
\end{figure}

\begin{figure}[c]
\centerline{\includegraphics[height=7cm,width=8cm]{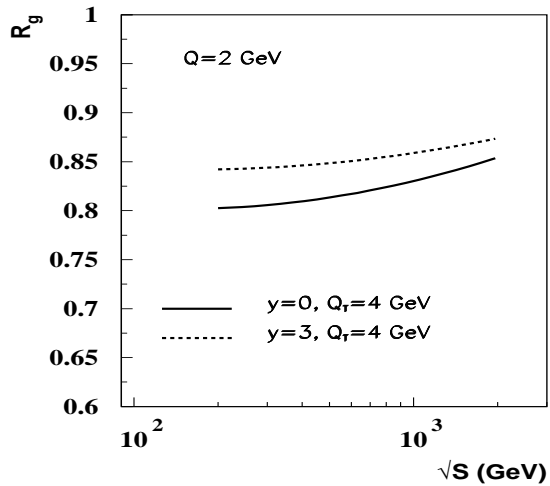}} 
\label{fig4}
\caption{The c.m. energy dependence of the ratio $R_{g}$ defined 
in Eq.~(\protect\ref{R-g}) for $Q=$2 GeV, $Q_T=$ 4~GeV, $y=0$ (solid) 
and $y=3$ (dashed). Note the logarithmic energy scale.}
\end{figure}
\section{The Drell-Yan cross section at RHIC and Tevatron energies}
\label{sec:X}

Recalling criterion (iii) in the Introduction, we now turn to 
the production rate of low-mass Drell-Yan pairs at 
RHIC and Tevatron.
We present the Drell-Yan cross section at large transverse
momentum with all-order resummation. For Drell-Yan pair production, as seen 
from Eq. (1), there is a phase space penalty associated with the finite mass of 
the virtual photon, and the Drell-Yan factor, $\alpha_{em}/(3\pi
Q^2)< 10^{-3}/Q^2$, renders the production rate for massive
lepton pairs small at large values of $Q$. In addition, the spectra 
drop rapidly with increasing $Q_T$. In order to enhance the
Drell-Yan cross section while keeping the dominance of the gluon
initiated sub-processes, it is useful to study lepton pairs with low
invariant mass and relatively large transverse momentum\cite{Berger:2001wr}.  
With the large transverse momentum $Q_T$ setting the hard scale of the 
collision, the invariant mass of the virtual photon $Q$ can be small,
as long as the process can be identified experimentally, and  
$Q\gg\Lambda_{\rm QCD}$.  For example, the cross section for Drell-Yan
production was measured by the CERN UA1 Collaboration\cite{UA1-Vph} 
for virtual photon mass $Q \in [2m_\mu, 2.5]$~GeV.  

Figure~5 shows the all-order resummed result
for the 
Drell-Yan cross section as a function of $Q_T$ at the RHIC
energy of $\sqrt{s}=200$ GeV and rapidity $y=0$ for two values of the mass, 
(a) $Q=1$ GeV and (b) $Q=2$ GeV. The cross section 
increases by about a factor
10 when the Drell-Yan mass decreases from 2 GeV to 1 GeV.
It might still be a challenge to 
measure the low-mass Drell-Yan production with the production rate shown
in Fig.~5. At the LHC, both the collision energy and luminosity are 
significantly improved. Our calculation shows that  
the production rate at the LHC is sufficiently large for being measured. 
(We also calculated $R_g$ for $Q=1$ GeV. 
When $Q$ decreases to 1 GeV, we find that $R_g$
actually increases slightly in the $Q_T\sim 1$ GeV region and 
it is similar to the $Q=2$~GeV case in the high $Q_T$ region.)

\begin{figure}
\begin{minipage}[c]{7.8cm}
\centerline{\includegraphics[width=8cm,height=8cm]{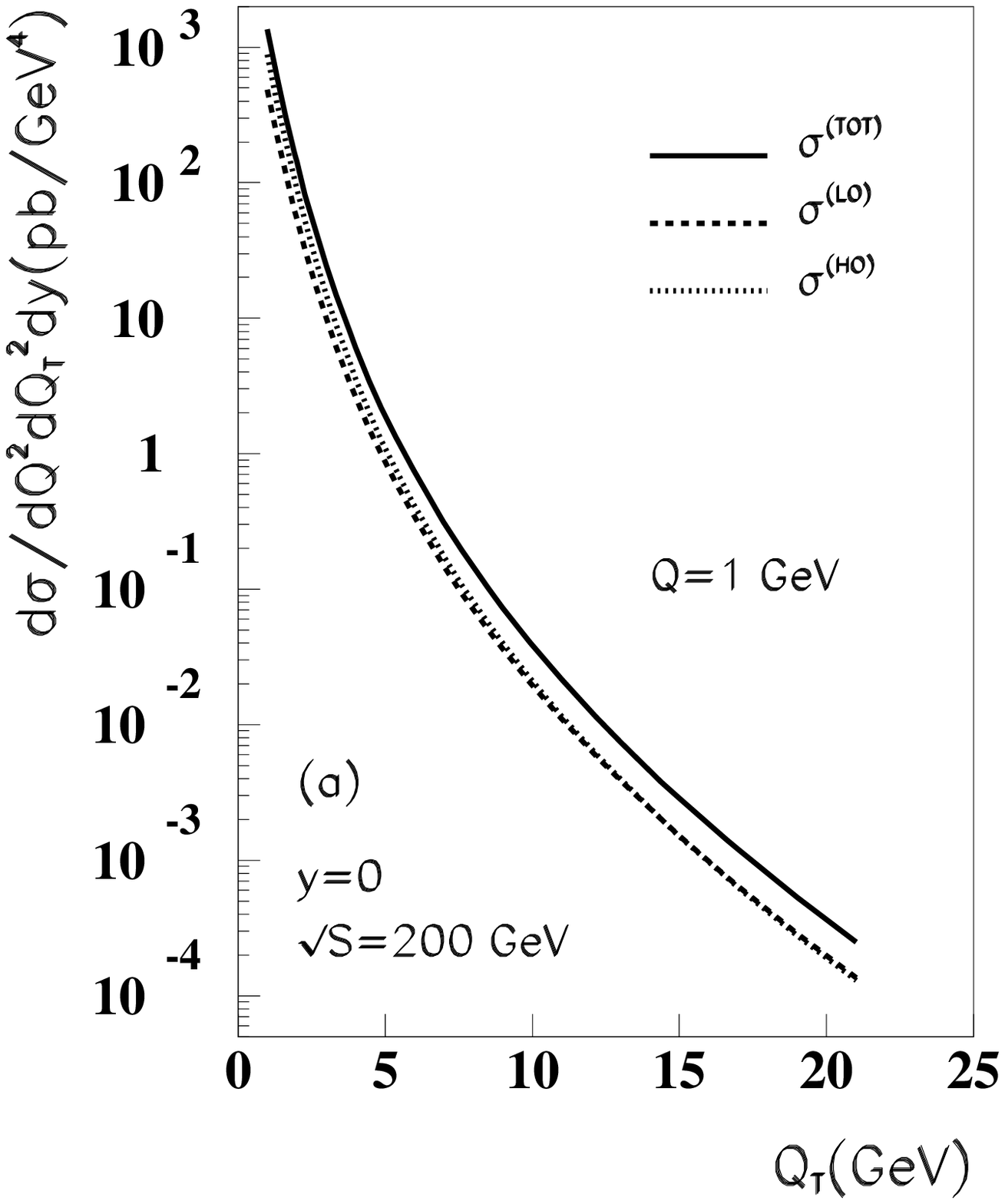}} 
\vspace{-0.1in}
\label{fig5}
\end{minipage}
\hfill
\begin{minipage}[c]{7.8cm}
\vspace{0.0in}
\includegraphics[width=8cm,height=8cm]{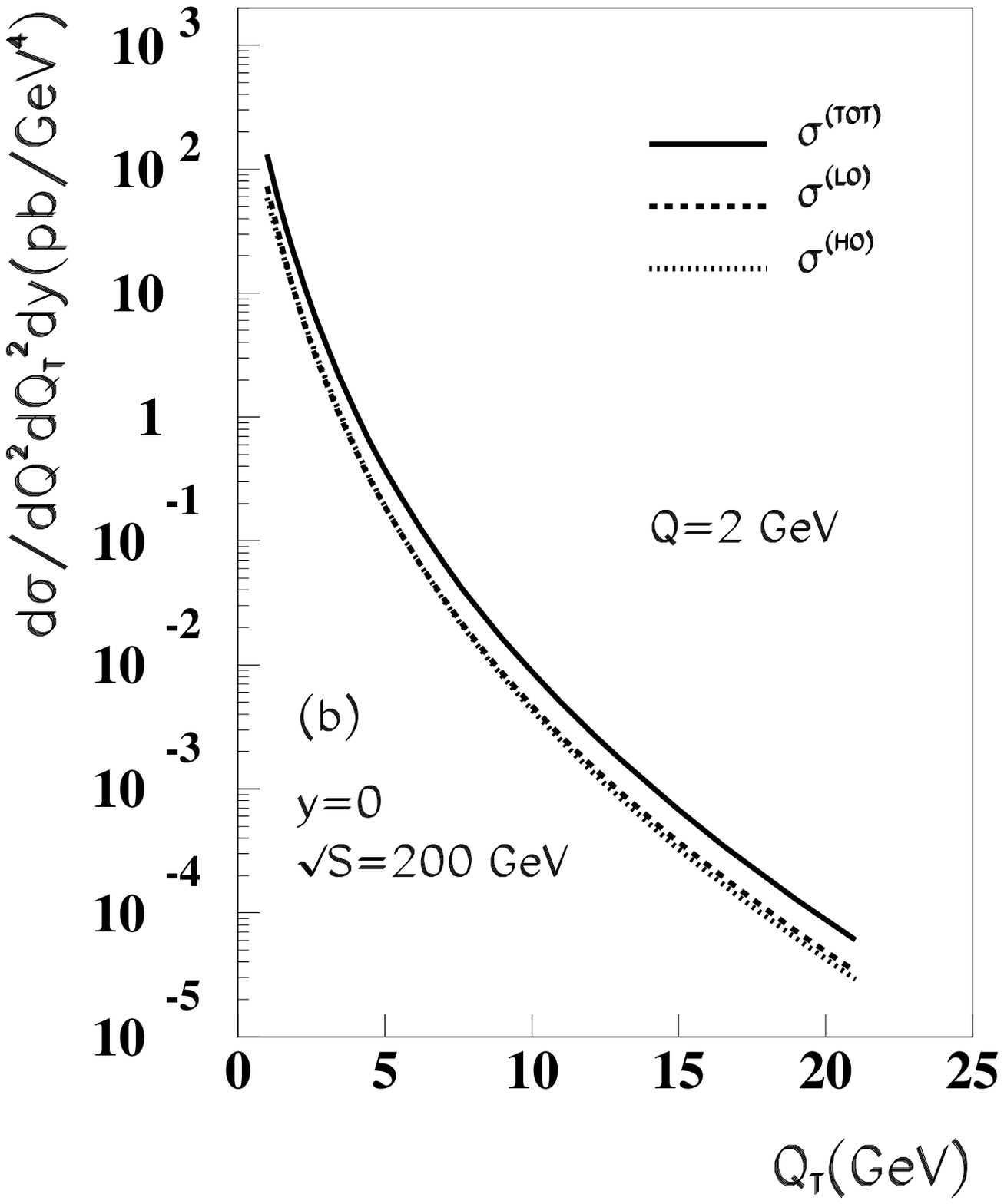} 
\vspace{-0.3in}
\end{minipage}
\vspace{0.2in}
\caption{Drell-Yan cross section as a function of $Q_T$ at RHIC
energy $\sqrt{s}=200$ GeV and rapidity $y=0$, for (a) mass 
$Q=1$ GeV, and (b) $Q=2$ GeV. The top (solid) lines represent
the total cross sections; leading-order (LO, dashed) and higher-order
(HO, dotted) contributions are also shown.
}
\end{figure}

Figure~6 displays similar results for the 
Tevatron
energies of  
$\sqrt{s}=1.96$~TeV ((a) and (b)) and $\sqrt{s}=630$~GeV (c).
For the larger energy, we show results with (a) $Q=5$~GeV  
and (b) $Q=2$~GeV. The cross section is about 20 times larger 
for $Q=2$~GeV than for $Q=5$~GeV at any fixed $Q_T$. 

In Figs.~5 and 6 we separately show the leading-order
(LO, dashed), and higher-order (HO, dotted) contributions. At 
high $Q_T$ the LO and HO contributions are roughly comparable. 
The total cross section can also be broken up into gluon-initiated, 
quark-quark, and resummed contributions. Results on this composition 
are published elsewhere\cite{Fai:2004qu}. 

\begin{figure}[tbh]
\vspace{0.5in}
\setlength{\unitlength}{.5in}
\begin{picture}(15,8)(.2,.5)
\label{fig6}
\put(-1.5, 0){\includegraphics{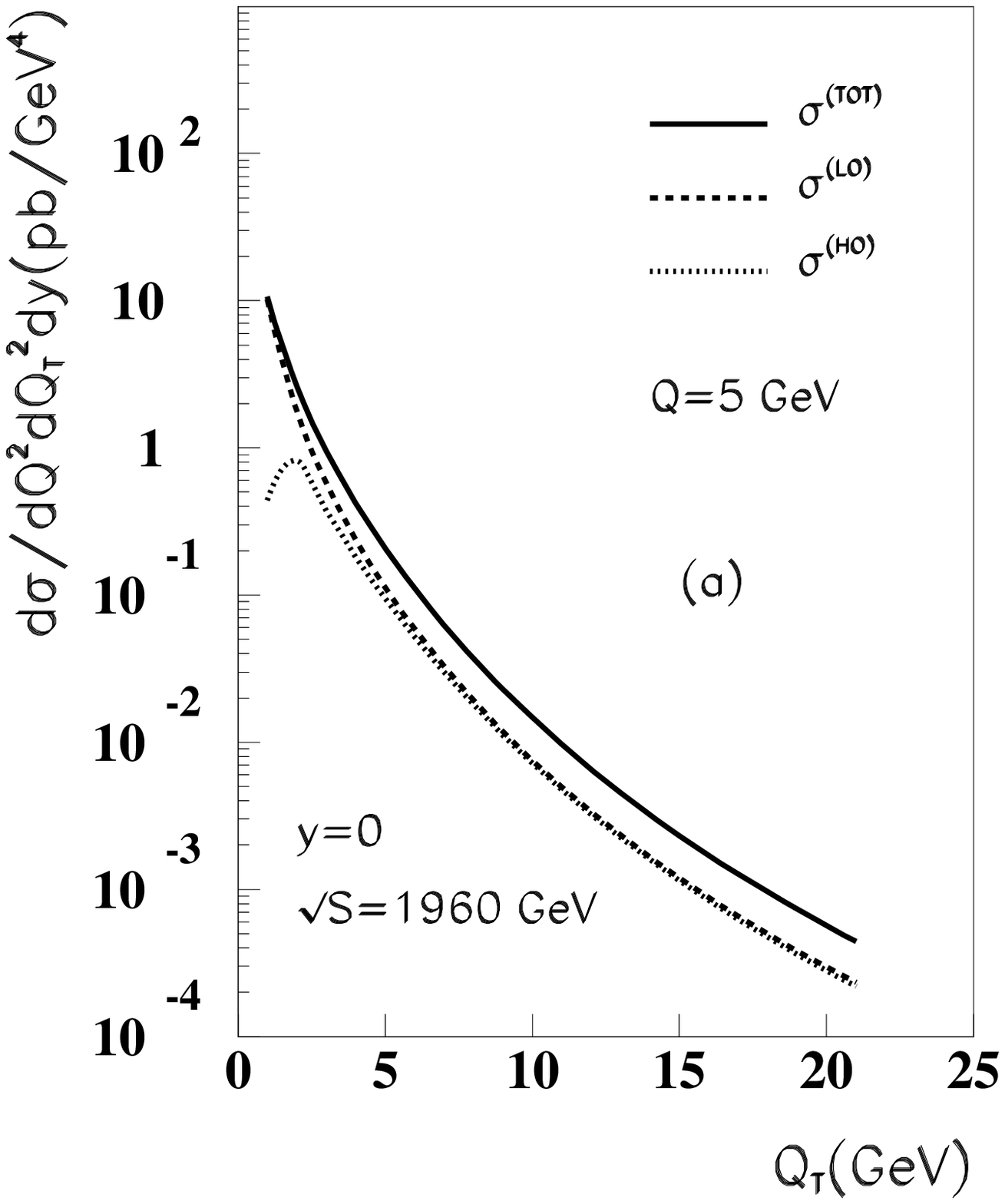}}
\put(3, 0){\includegraphics{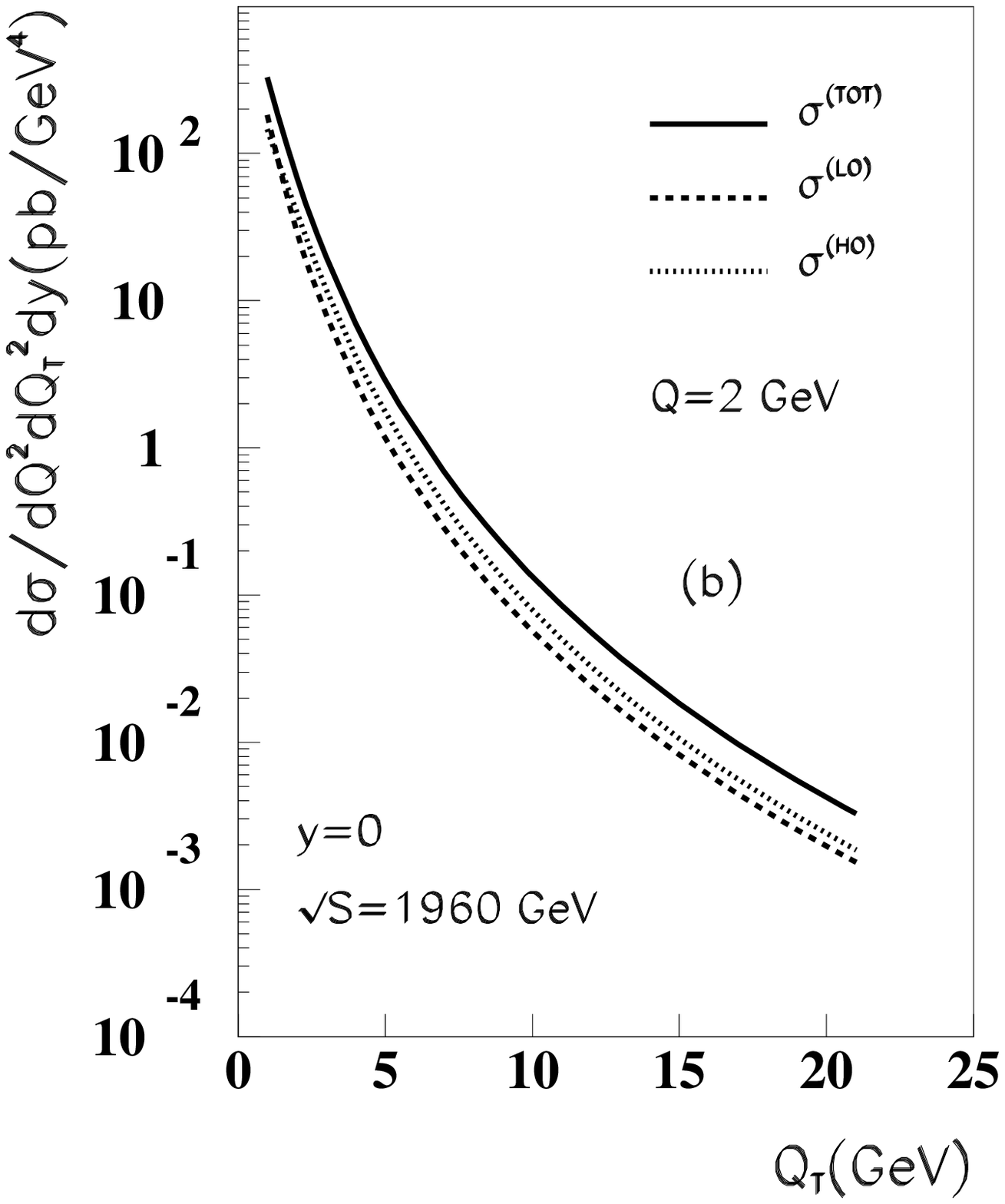}}
\put(7.5, 0){\includegraphics{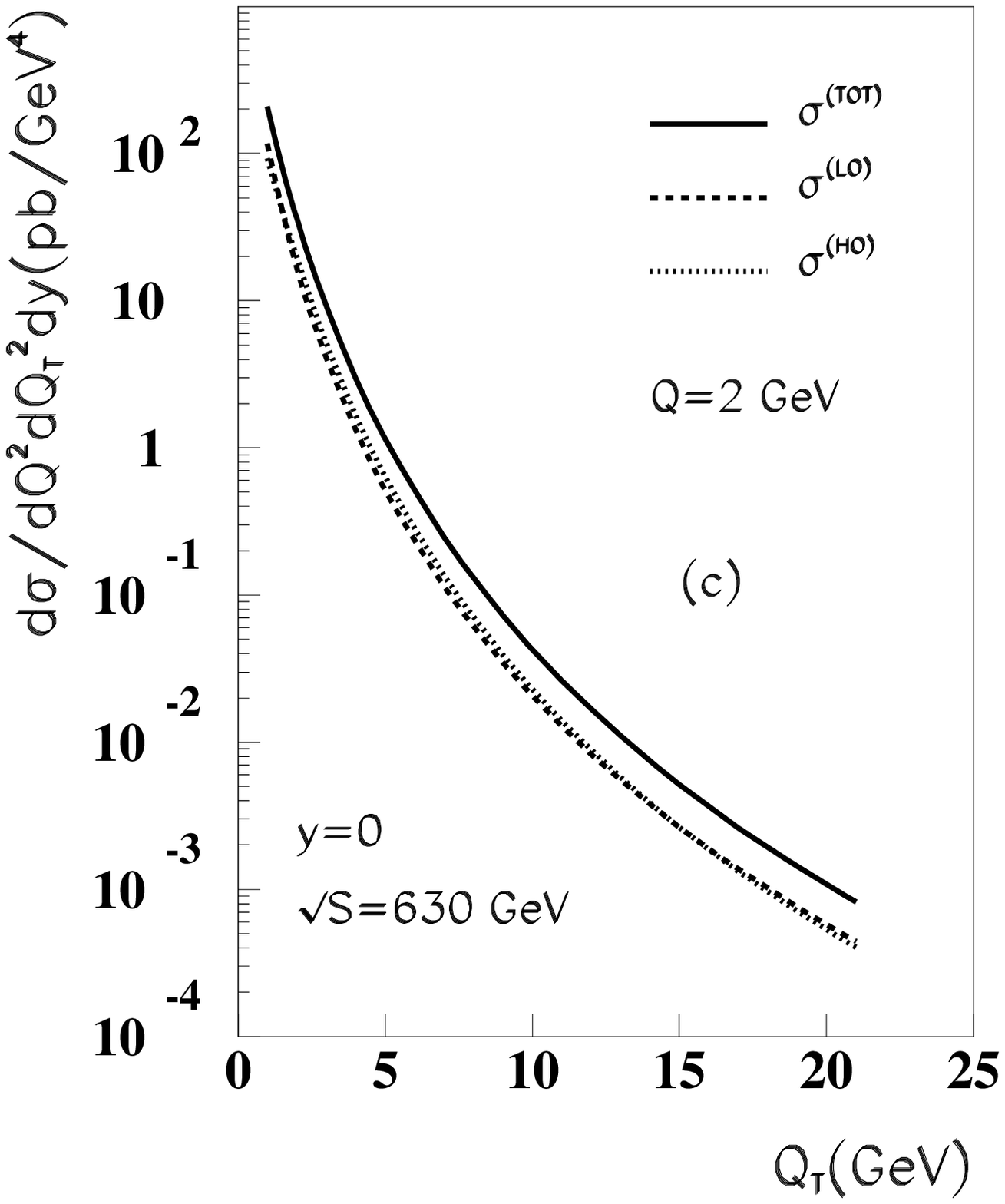}}
\end{picture}
\vspace*{-1.0in}
\caption{Drell-Yan cross section at rapidity $y=0$ as a function of $Q_T$ 
at 
Tevatron
energies: (a) $\sqrt{s}=1.96$~TeV and $Q=5$~GeV; 
(b) $\sqrt{s}=1.96$~TeV and $Q=2$~GeV; and (c) $\sqrt{s}=630$~GeV and 
$Q=2$~GeV. The top (solid) lines represent the total cross sections; 
leading-order (LO, dashed) and higher-order (HO, dotted) contributions 
are also shown.
\vspace*{0.4in}
}
\end{figure}

Figures~7 and 8 present the rapidity dependence 
of the cross sections at RHIC and 
Tevatron
energies at various values of 
$Q$ and $Q_T$. It can be seen that while there are order-of-magnitude
differences in the cross section between different values of $Q$ and 
$Q_T$, the rapidity distributions remain flat upto high absolute
values of $y$. This lends further support to the idea of exploring
forward rapidities. As we will see in Section \ref{sec:xrange}, the 
forward region is an excellent place to probe small-$x$ gluons.

\begin{figure}

\begin{minipage}[c]{7.8cm}
\centerline{\includegraphics[width=8cm,height=8cm]{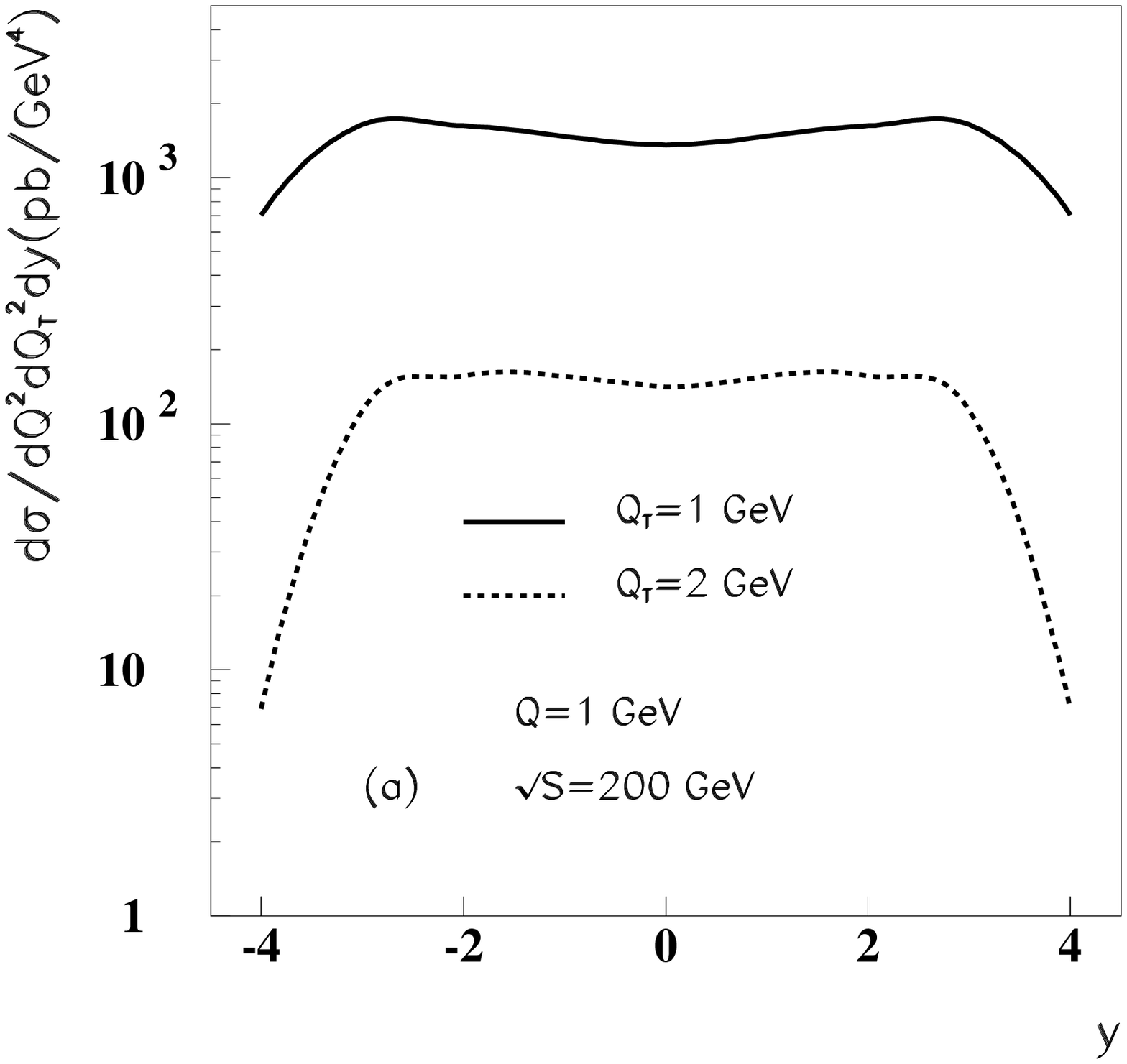}} 
\vspace{-0.1in}
\label{fig7}
\end{minipage}
\hfill
\begin{minipage}[c]{7.8cm}
\vspace{0.0in}
\includegraphics[width=8cm,height=8cm]{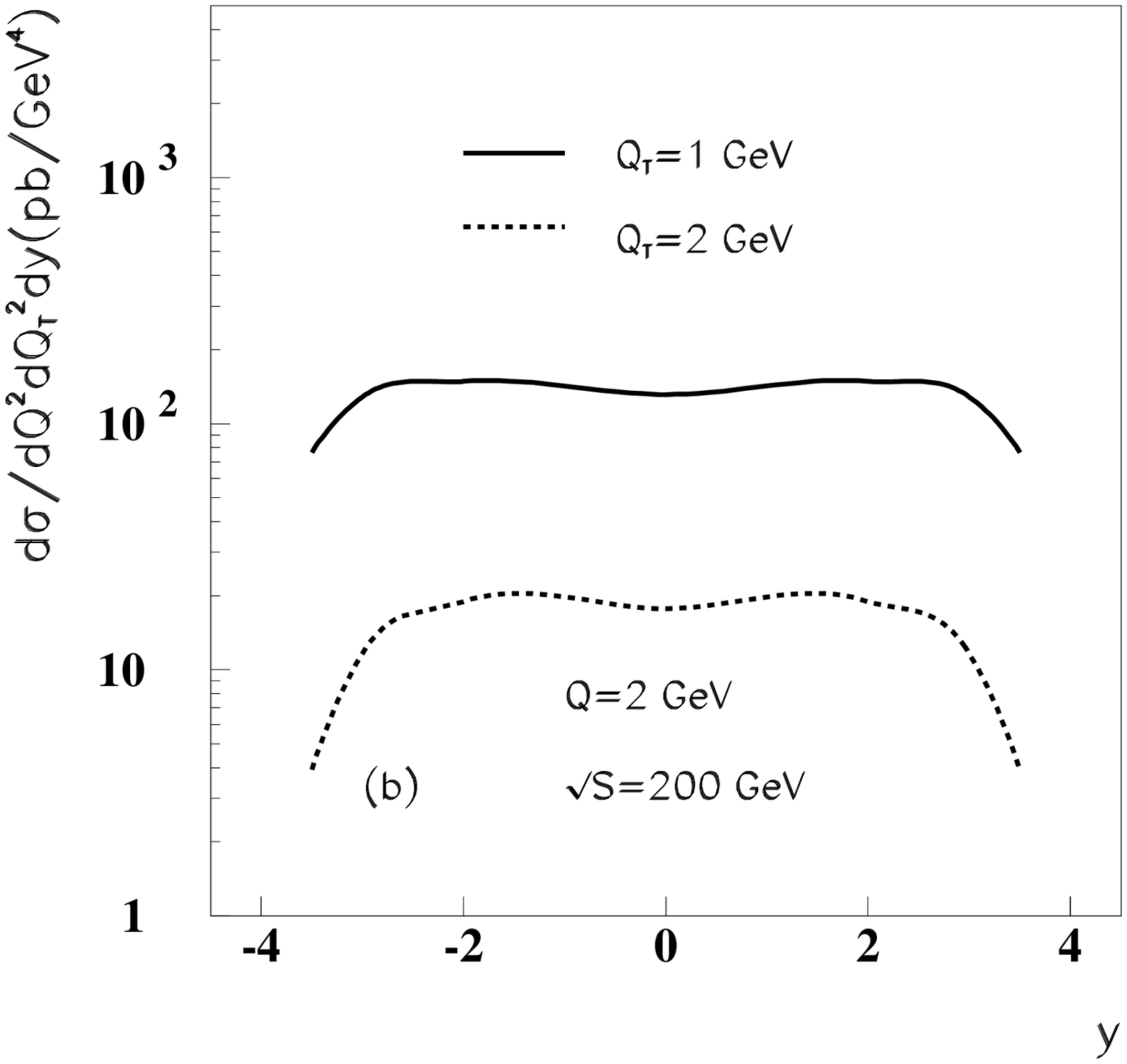} 
\vspace{-0.3in}
\end{minipage}
\vspace{0.2in}
\caption{Drell-Yan cross section as a function of $y$ at RHIC
energy $\sqrt{s}=200$ GeV with $Q_T=1$~GeV (solid) and 
$Q_T=2$~GeV (dashed) for (a) $Q=1$~GeV and (b) $Q=2$~GeV.
}
\end{figure}

\begin{figure}
\begin{minipage}[c]{7.8cm}
\centerline{\includegraphics[width=8cm,height=8cm]{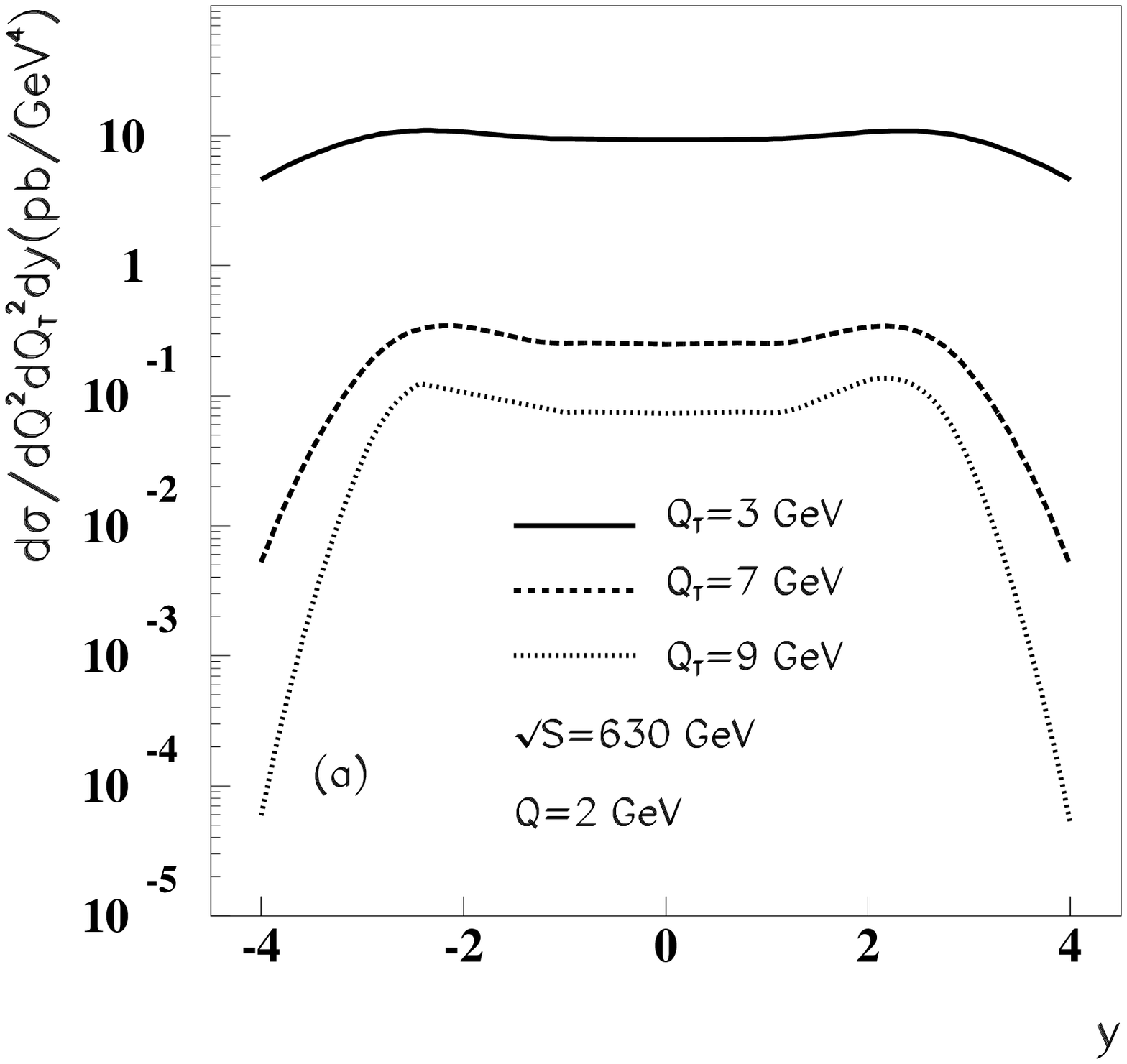}} 
\vspace{-0.1in}
\label{fig8}
\end{minipage}
\hfill
\begin{minipage}[c]{7.8cm}
\vspace{0.0in}
\includegraphics[width=8cm,height=8cm]{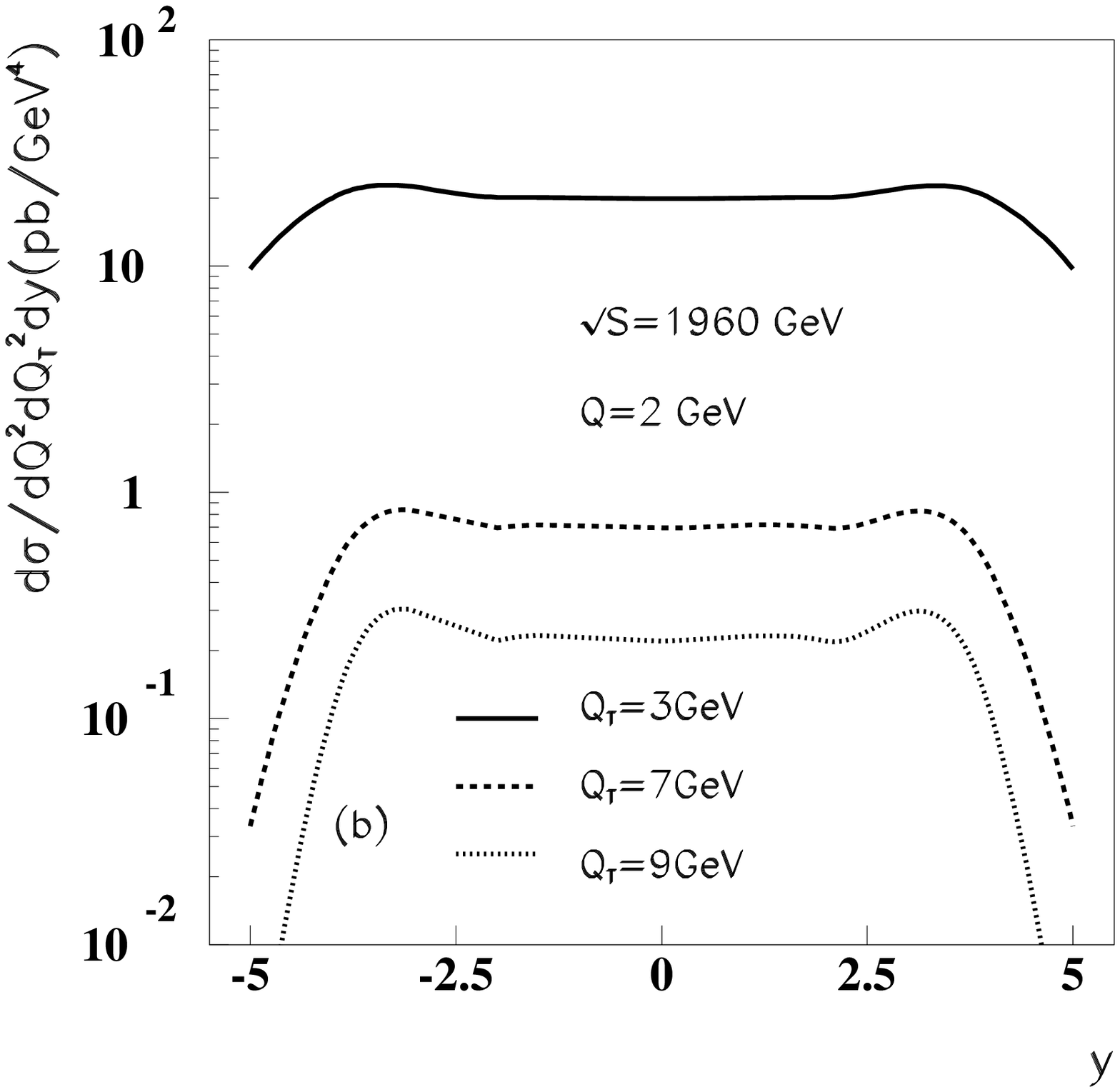} 
\vspace{-0.3in}
\end{minipage}
\vspace{0.3in}
\caption{Drell-Yan cross section as a function of rapidity $y$
for $Q=2$~GeV and $Q_T=3$~GeV (solid), 7~GeV (dashed),
and 9~GeV (dotted) at 
Tevatron
energies: (a) $\sqrt{s}=630$~GeV,
and (b) $\sqrt s=1960$~GeV.
}
\end{figure}

\section{Parton momentum fraction probed}
\label{sec:xrange}

It is very important to estimate the region of $x$ in the gluon  
distribution probed by the low-mass Drell-Yan process. Let us use 
$x_1$ to refer to partons in one of the beams (say the ``yellow'' 
beam in the RHIC color code), and $x_2$ for partons in the other
(``blue'') beam. The $x_1$ and $x_2$ integrations of the Dell-Yan 
cross section run from the appropriate minimum values
(given by the kinematics) to 1. At central rapidity $y=0$, 
the integration is symmetric in $x_1$ and $x_2$. 
We are most interested in small $x$ physics. 
 We expect a large fraction of the cross section to come 
from rather small values of $x_2$ in the forward region
from the perspective of the yellow beam, where $x_1$ is not small.
To quantify  our statement with respect to $x_2$, we 
introduce a cutoff $x_{cut}$ to limit the $x_2$ integration to 
the interval $x_{min}$ to $x_{cut}$ for partons in the blue beam. 
Let us define the ratio
\begin{equation}
R_{x_2} = \left.
          \int_{x_{\rm min}}^{x_{\rm cut}} dx_2
          \left( \frac{d\sigma^{\rm DY}}{dx_2} \right)
          \right/
          \int_{x_{\rm min}}^{1} dx_2
          \left( \frac{d\sigma^{\rm DY}}{dx_2} \right)  \,\,\, .
\label{Rx2}
\end{equation}
Due to the symmetry in $x_1$ and $x_2$, at central rapidities it is enough 
to examine the shape of $R_{x_2}$ to establish which region 
dominates the $x$ integration --- the region of $x$ low-mass Drell-Yan data 
can provide precise information about. In Fig.~9(a) we plot $R_{x_2}$ at 
RHIC, as a function of $x_{cut}$ for $Q=1$ GeV 
for different rapidities and transverse momenta. Figure~9(a) shows that
in the central rapidity region $x_2\sim [10^{-2}, 10^{-1}]$
dominates the integration for both $Q_T=1$~GeV and $Q_T=3$ GeV when $y=0$
at RHIC.

Figure 9(a) also provides important information about the forward region
(from the perspective of $x_1$). At $y=3$, 90\% of the cross-section is given 
by $x_2 < 0.01$, which is exactly the shadowing region of 
the nuclear parton distribution function. 
This region dominates the integration for both $Q_T=1$ GeV and $Q_T=3$ GeV 
when $y=3$. Our calculation for the rapidity distribution of the total cross 
section (see Fig.~7) shows that the cross section does not start to drop until 
$y\sim 3$. In other words, the production rate in the large-rapidity region
is not significantly smaller than in the central region, as long as 
$|y|\le 3$.
 While, as indicated by Fig.~9(a), the typical $x_2$ is much smaller 
in the forward region than at $y=0$, the situation is different for $x_1$.
In order to evaluate the important region of $x_1$ for Drell-Yan production, 
we define the analogous ratio (compare to Eq. (\ref{Rx2}))
\begin{equation}
R_{x_1} = \left.
          \int_{x_{\rm min}}^{x_{\rm cut}} dx_1
          \left( \frac{d\sigma^{\rm DY}}{dx_1} \right)
          \right/
          \int_{x_{\rm min}}^{1} dx_1
          \left( \frac{d\sigma^{\rm DY}}{dx_1} \right)  \,\,\, .
\label{Rx1}
\end{equation}

In Fig. 9(b), we plot $R_{x_1}$ for forward rapidity $y=3$ and $Q=$1~GeV 
at $Q_T=1$ GeV  (solid) and $Q_T=3$~GeV  (dashed). 
For $Q_T=1$ GeV, the dominant region is 
$x_1\sim [0.1,0.3]$ and for $Q_T=3$ GeV, the dominant region is 
$x_1\sim [0.3, 0.7]$. (The ratio $R_{x_1}$ for $y=0$ is of course 
identical to $R_{x_2}$ for $y=0$ due to the symmetry, and thus 
can be seen in Fig.~9(a).) As we will discuss later, the dominant $x_2$ 
region plays an important role in the nuclear effects on Drell-Yan 
production in an $AB$ nuclear collision. 

\begin{figure}
\begin{minipage}[c]{7.8cm}
\centerline{\includegraphics[width=8cm,height=7cm]{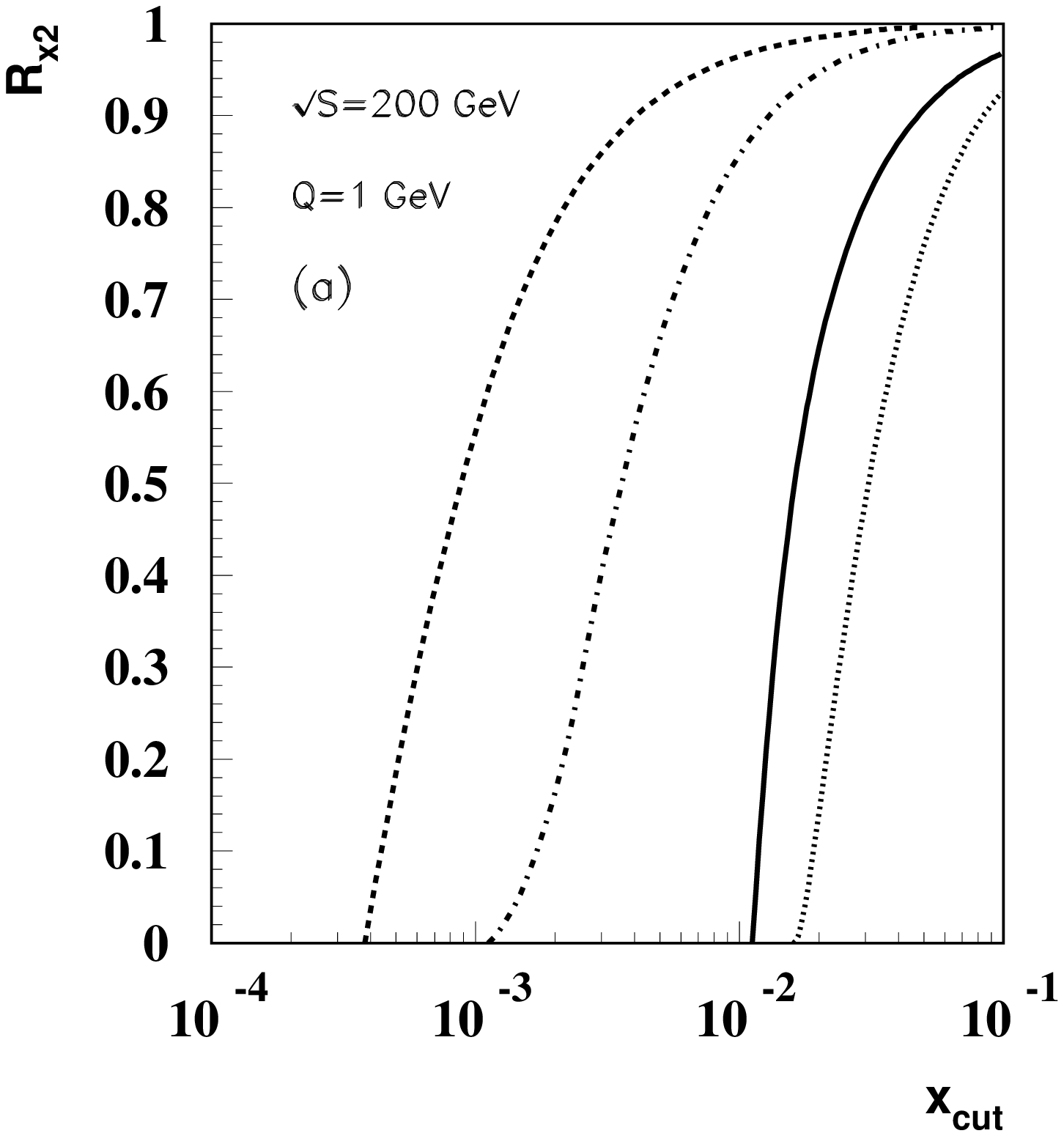}} 
\label{fig9}
\end{minipage}
\hfill
\begin{minipage}[c]{7.8cm}
\includegraphics[width=8cm,height=7cm]{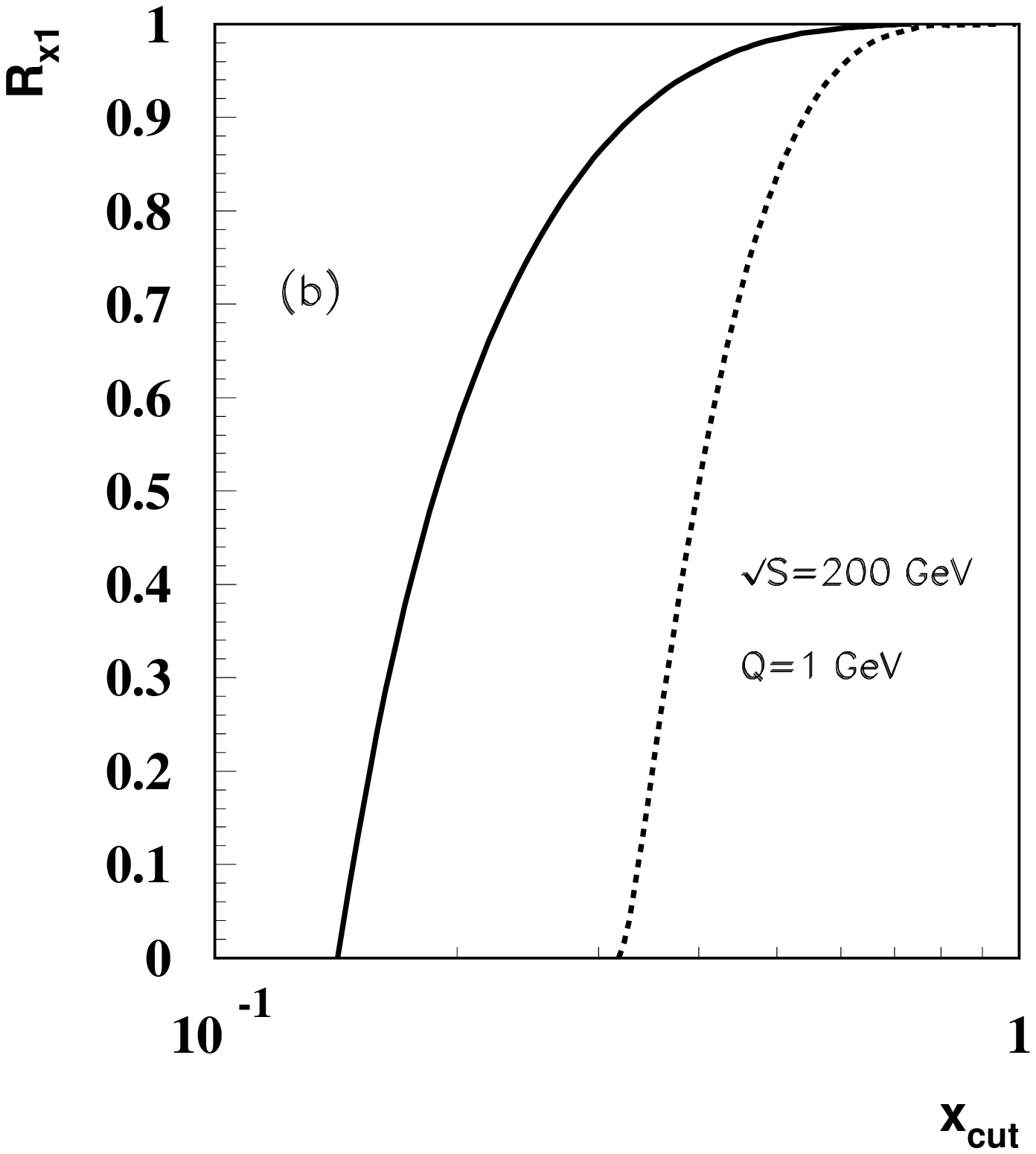} 
\end{minipage}
\vspace{0.2in}
\caption{The ratios $R_{x2}$ (defined in 
Eq.~(\protect\ref{Rx2})) and  $R_{x1}$ (Eq.~(\protect\ref{Rx1}))
for $Q=1$ GeV at RHIC energy, $\sqrt s=200$ GeV. In (a),
both central rapidities: $y=0, Q_T=1$ GeV (solid); $y=0, Q_T=3$ GeV 
(dotted), and forward rapidities: $y=3, Q_T=1$ GeV (dashed); $y=3, 
Q_T=3$ GeV (dot-dashed) are shown. In (b) $R_{x1}$ is displayed 
at forward rapidities: $y=3, Q_T=1$ GeV (solid); 
$y=3, Q_T=3$~GeV (dashed). See text for more explanation.}
\end{figure}

Figures 10 and 11 are similar to Fig.~9, but for the top 
Tevatron
energy of 
$\sqrt{s}=$1.96~TeV and for the LHC energy of 5.5~TeV, respectively.
Fig.~10(a) displays $R_{x_2}$ as a function of $x_{cut}$ for $Q=2$ GeV 
for different rapidities and transverse momenta $Q_T$. Fig.~10(b)
shows $R_{x_1}$ for $y=3$ at $Q_T=3$ GeV and $Q_T=7$ GeV.
 It can be seen from Fig.~10(a) that the dominant region in $x_2$ 
at central rapidity is $x_2\sim [10^{-3}, 10^{-2}]$ at the top 
Tevatron
energy;
at forward rapidities $x_2 < 10^{-4}$ is reached. For $x_1$ in the forward
region, the dominant contribution comes from $x_1 \sim [4*10^{-2},10^{-1}]$,
as seen in Fig.~10(b).
In Fig.~11(a) we plot $R_{x_2}$ as a function of $x_{cut}$ for $Q=2$ GeV 
for different rapidities and transverse momenta at the LHC. Figure~11(a) 
shows that in the central rapidity region, even when $Q_T=10$ GeV, $x_2<0.01$, 
i.e. the shadowing region, dominates the contribution to the Drell-Yan cross 
section. In the forward rapidity region with $y=2.4$, low-mass Drell-Yan 
production will probe partons with $x_2$ as small as on the order of $10^{-5}$. 
In Fig.~11(b), we plot $R_{x_1}$ for $y=2.4$ at $Q_T=2$~GeV 
and $Q_T=10$ GeV. The information provided here 
will be used in the next section (Section \ref{sec:nucl})
to discuss the nuclear effects in the forward region at the LHC.

\begin{figure}
\begin{minipage}[c]{7.8cm}
\centerline{\includegraphics[width=8cm,height=7cm]{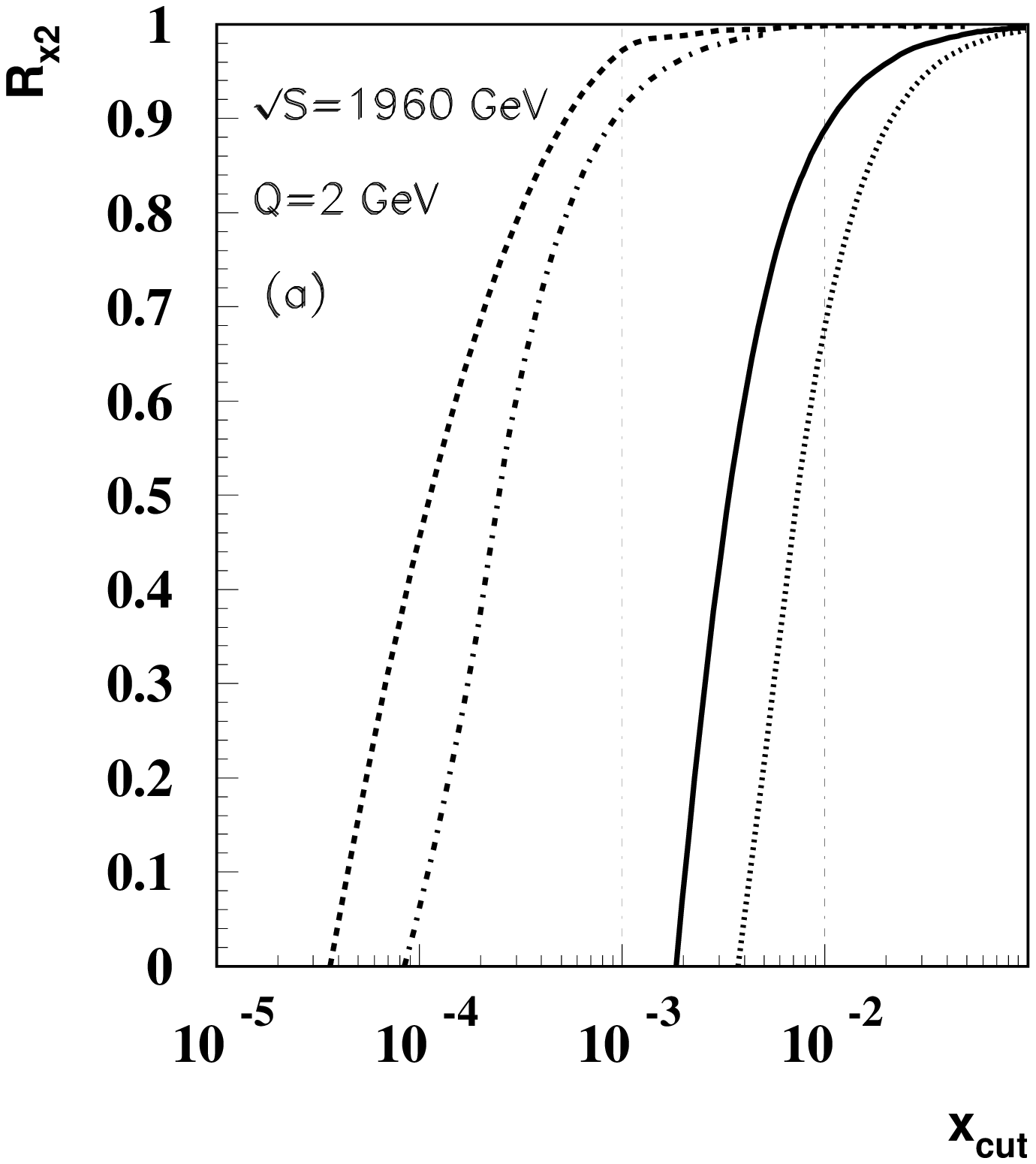}} 
\label{fig10}
\end{minipage}
\hfill
\begin{minipage}[c]{7.8cm}
\includegraphics[width=8cm,height=7cm]{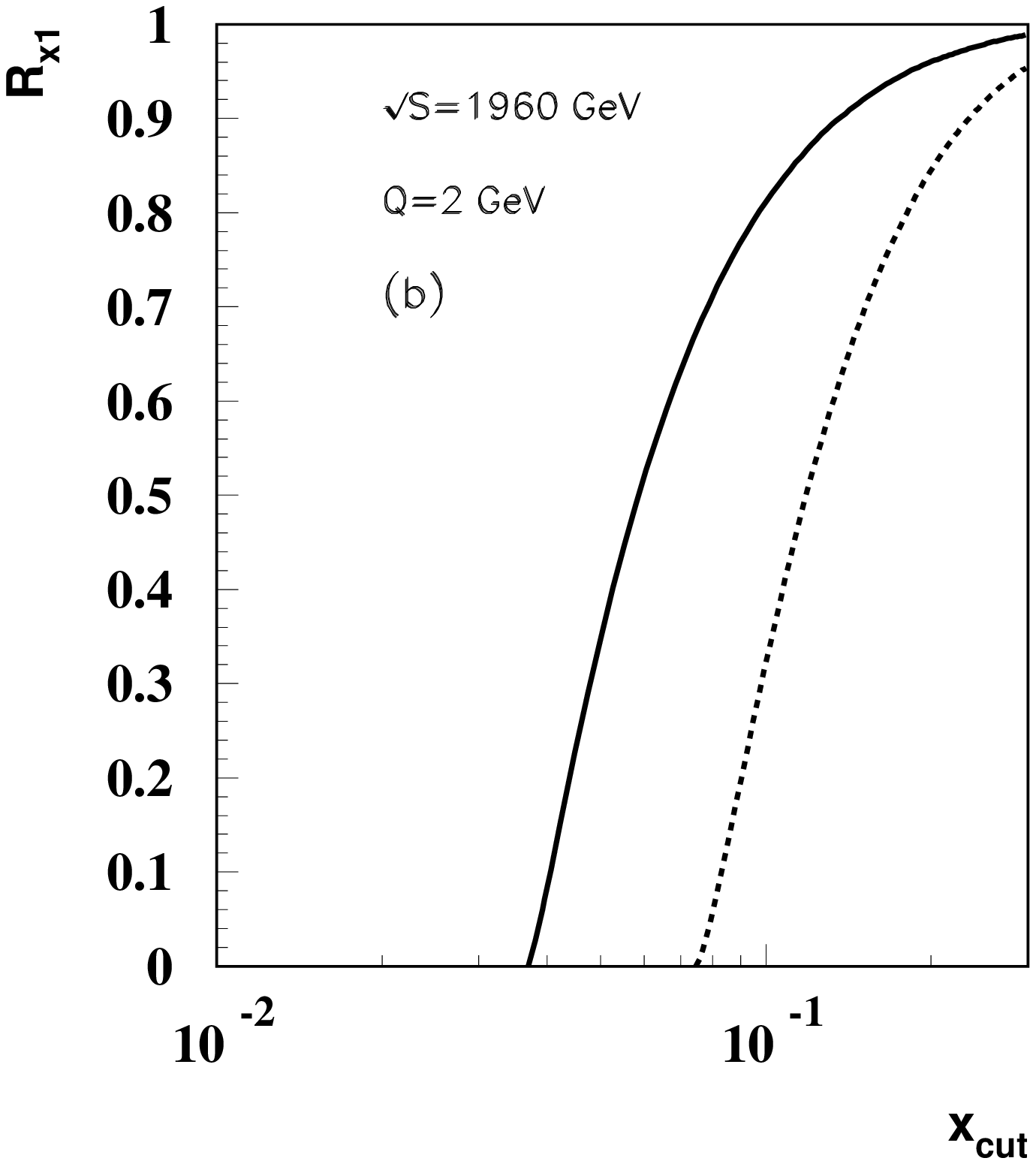} 
\end{minipage}
\vspace{0.2in}
\caption{The ratios $R_{x2}$ (defined in 
Eq.~(\protect\ref{Rx2})) and  $R_{x1}$ (Eq.~(\protect\ref{Rx1}))
for $Q=2$ GeV at 
Tevatron
energy, $\sqrt s=1.96$ TeV. In (a),
both central rapidities: $y=0, Q_T=3$ GeV (solid); $y=0, Q_T=7$ GeV 
(dotted), and forward rapidities: $y=3, Q_T=3$ GeV (dashed); $y=3, 
Q_T=7$ GeV (dot-dashed) are shown. In (b) $R_{x1}$ is displayed 
at forward rapidities: $y=3, Q_T=3$ GeV (solid); 
$y=3, Q_T=7$~GeV (dashed).}
\end{figure}

\begin{figure}
\begin{minipage}[c]{7.8cm}
\centerline{\includegraphics[width=8cm,height=7cm]{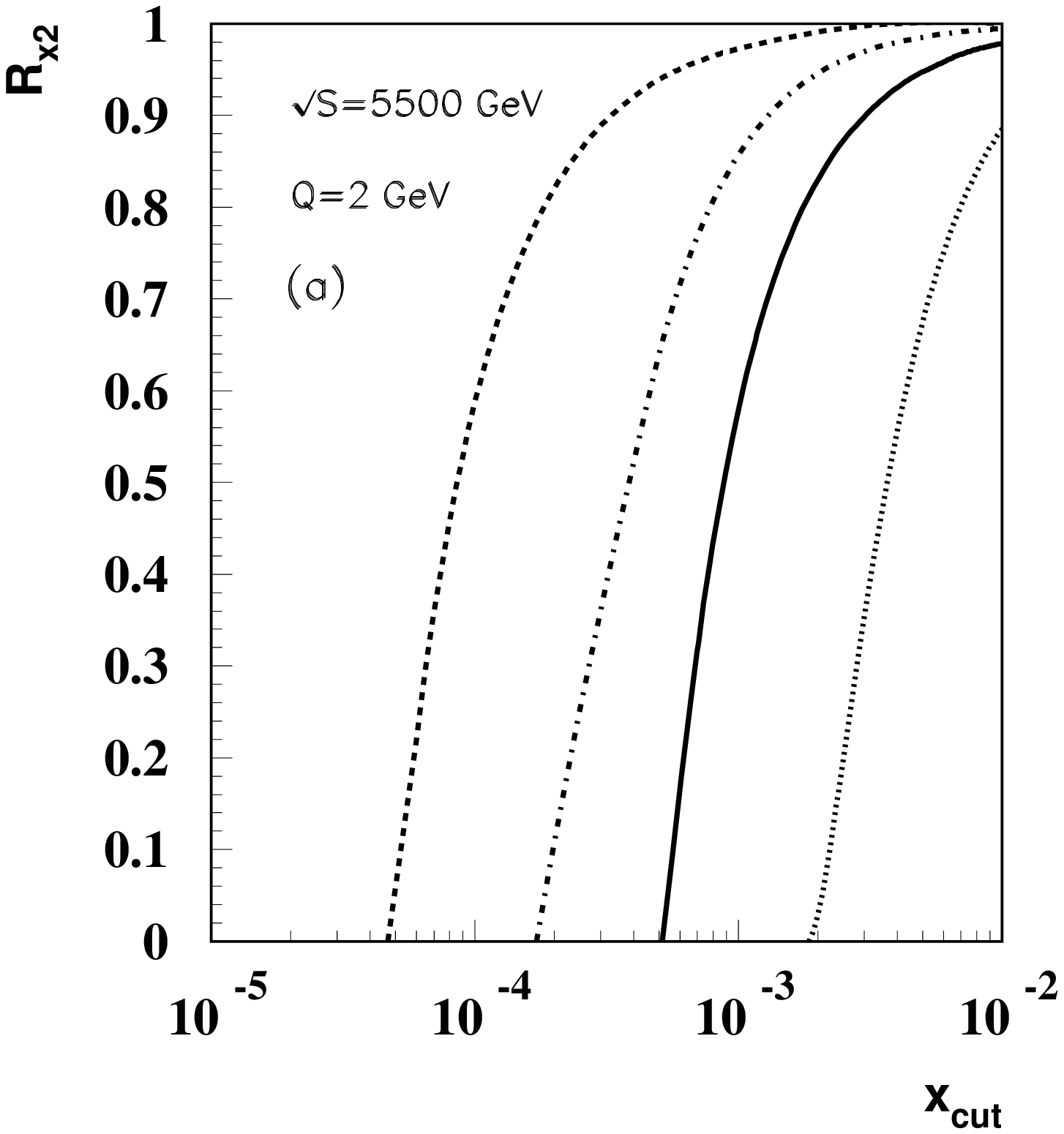}} 
\label{fig11}
\end{minipage}
\hfill
\begin{minipage}[c]{7.8cm}
\includegraphics[width=8cm,height=7cm]{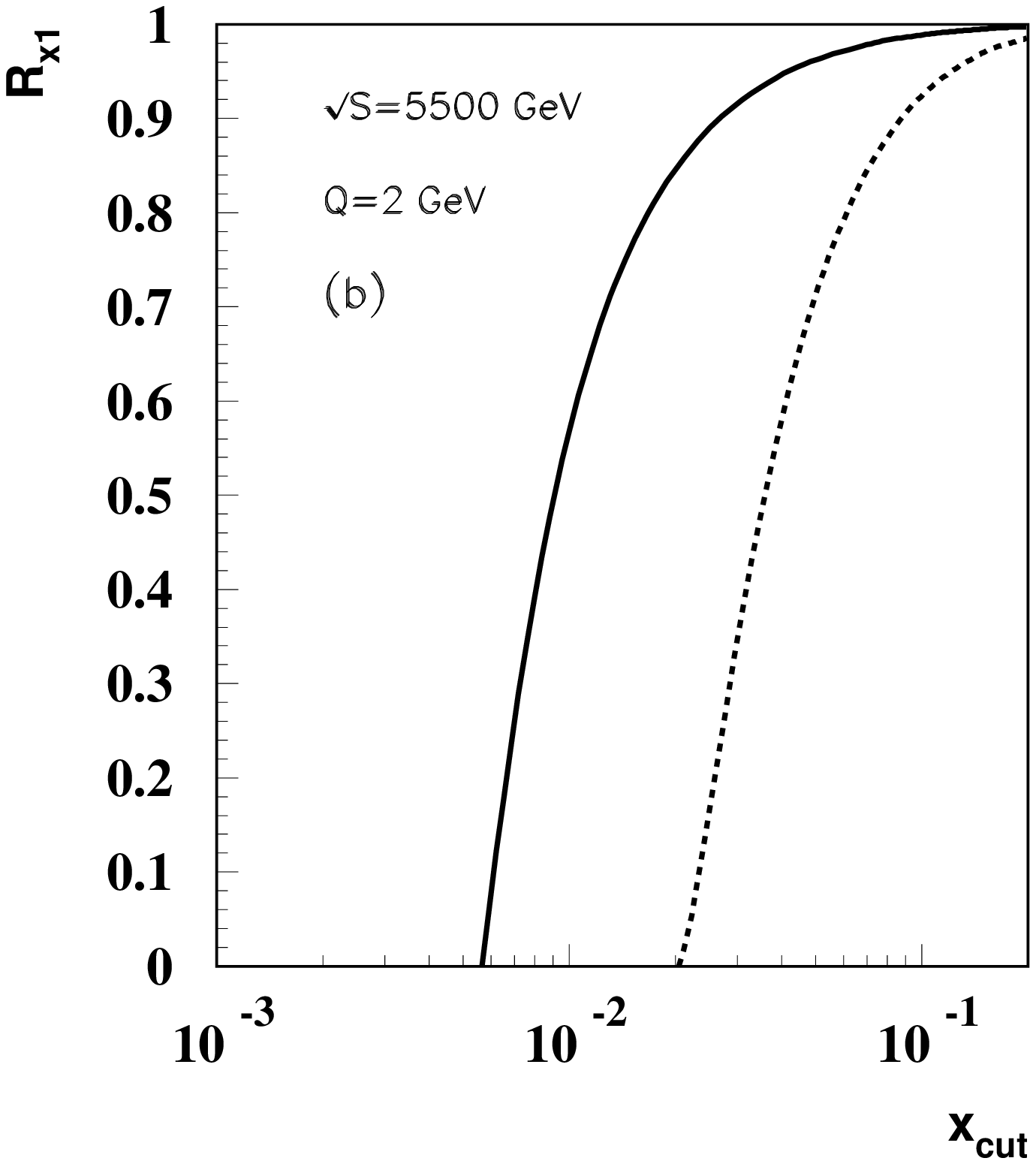} 
\end{minipage}
\vspace{0.2in}
\caption{The ratios (a) $R_{x_{2}}$ and (b) $R_{x_{1}}$
for $Q=2$ GeV at LHC $\sqrt s=5500$ GeV.
Different lines in (a) represent $R_{x2}$ at different $y$
and $Q_T$: $y=0, Q_T=2$ GeV (solid); $y=0, Q_T=10$ GeV (dotted);
$y=2.4, Q_T=2$ GeV (dashed); $y=2.4, Q_T=10$ GeV (dot-dashed).
In (b) we have $y=2.4, Q_{T}=2$ GeV (solid) and 
$y=2.4, Q_{T}=10$ GeV (dashed).}
\end{figure}

\section{Nuclear effects at RHIC and LHC}
\label{sec:nucl}

In lack of nuclear effects on the hard collision,
the impact-parameter integrated production cross section
of low-mass Drell-Yan pairs in nucleus-nucleus ($AB$) 
collisions should scale, compared to the production in $pp$ collisions,
as the number of 
binary 
collisions.
However, there are several nuclear effects on the hard collision in a
heavy-ion reaction. For high transverse-momentum Drell-Yan production,   
two kinds of nuclear effects are expected to play an important role:
isospin effects and the modification of the parton distribution function 
in the nucleus (shadowing effect). In this Section, we calculate 
 a ``nuclear modification factor'' $R_{AB}$
at RHIC and LHC, and discuss isospin and shadowing effects 
in Drell-Yan production. We consider different rapidities, and for the
RHIC program, compare $AuAu$ and $dAu$ collisions. 
Isospin effects were recently also emphasized in
$dAu$ scattering at forward rapidities in Ref.~\cite{Guzey:2004zp}.

Similarly to the ``nuclear modification factor'' in hadron production, 
we define a ratio,
\begin{equation}
R_{AB}(Q_T) \equiv \left.
\frac{d\sigma^{(sh)}(Z_A/A,Z_B/B)}{dQ_T^2 dy} \right/
\frac{d\sigma}{dQ_T^2 dy} \,\,\, ,
\label{rab}
\end{equation}
where $Z_A$ and $Z_B$ are the atomic numbers and $A$ and $B$ are 
the mass numbers of the colliding nuclei, and the cross section
$d\sigma^{(sh)}(Z_A/A,Z_B/B)/dQ_T^2 dy$  uses nuclear PDFs 
defined in Eq.~(\ref{iso}) below,
while $d\sigma/dQ_T^2 dy$ is the $pp$ cross section
defined in Eq.~(\ref{DY-fac}).
Thus, $R_{AB}$ incorporates the effects of both shadowing and  
nuclear composition (isospin effect).

\begin{figure}
\begin{minipage}[c]{7.8cm}
\centerline{\includegraphics[width=8cm,height=7cm]{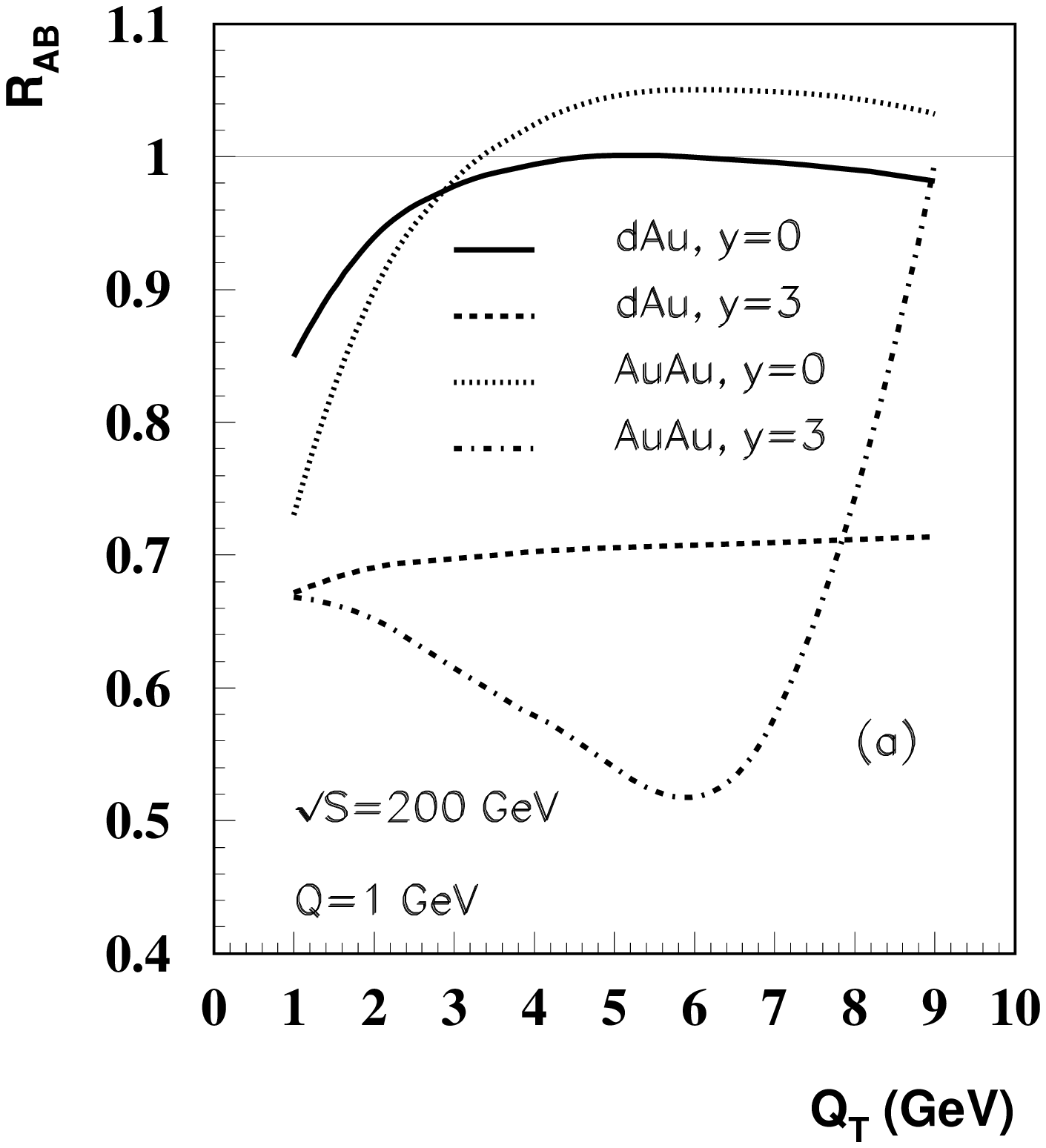}} 
\label{fig12}
\end{minipage}
\hfill
\begin{minipage}[c]{7.8cm}
\includegraphics[width=8cm,height=7cm]{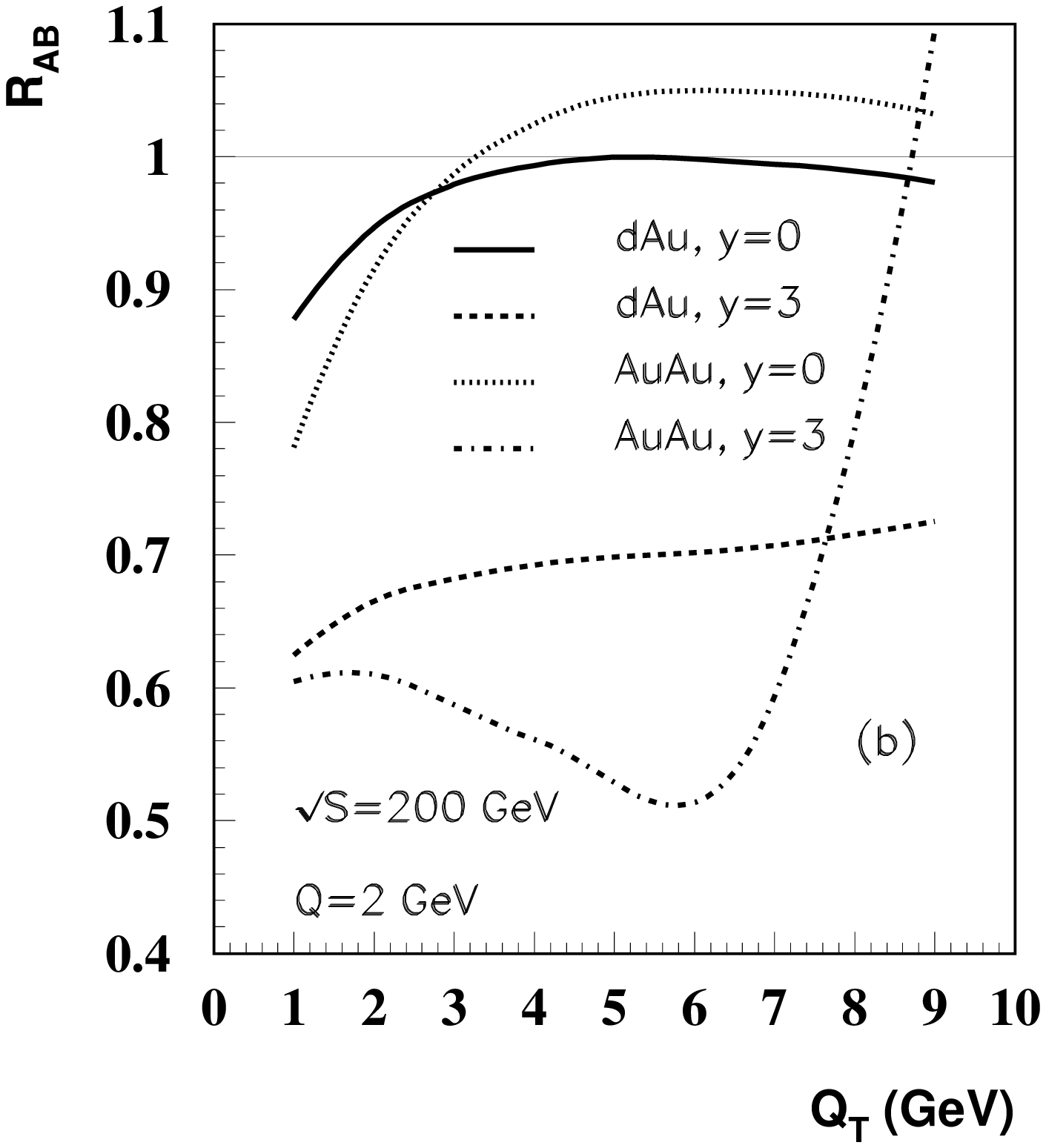} 
\end{minipage}
\vspace{0.2in}
\caption{The ratio $R_{AB}$ defined in 
Eq.~(\protect\ref{rab})
for (a) $Q=1$ GeV and (b) $Q=2$ GeV. 
Different lines in the figure represent $R_{AB}$ at different 
values of $y$ for different collisions.}
\end{figure}

In Fig.~12, we plot $R_{AB}$ as a function of $Q_T$ at RHIC in $dAu$
and $AuAu$ collisions at both central rapidity ($y=0$) and 
forward rapidity ($y=3$) for (a) $Q=1$ GeV and (b) $Q=2$~GeV.  In this work
we use the EKS parameterization to incorporate the shadowing effect in the 
nuclear PDF \cite{Eskola:1998df}.  
Numerical differences from using other nuclear PDF  parameterizations
\cite{Qiu:1986wh,Hirai:2001np,Huang:1997ii,Li:2001xa,Frankfurt:2002kd,Armesto:2002ny}
could be estimated from the ranges of momentum fraction $x$ probed at
different collision energies. Fig.~12 shows that
in the central rapidity region, $R_{dAu}$ is
larger than $R_{AuAu}$ in the small $Q_T$ region, and $R_{AuAu}$ becomes 
larger than one and larger than $R_{dAu}$ when $Q_T$ is larger than $\approx 3$~GeV.
This seems to be surprising at first glance. Fig.~12 also shows strong suppression of 
Drell-Yan pair production in both $dAu$ and $AuAu$ collisions in the forward region. 
Furthermore, $R_{AuAu}$ increases steeply at large $Q_T >$ 6~GeV when $y=3$.
In order to disentangle shadowing and isospin effects, both incorporated in 
$R_{AB}$ as defined in Eq.(\ref{rab}), we separate these dependences and
investigate isospin and shadowing effects separately in the following 
discussion.

To take into account the neutron-proton composition of nuclei
one normally uses 
\begin{equation}
\Phi_{a/A}=\frac{Z_{A}}{A}\Phi_{a/A}^p + (1-\frac{Z_{A}}{A})\Phi_{a/A}^n
\label{iso}
\end{equation}
for the nuclear PDF, where $\Phi_{a/A}^p$ is the proton PDF
and $\Phi_{a/A}^n$ is the PDF of the neutron. By artificially
setting $Z_A=A$ and $Z_B=B$ in our calculation, we consider 
nuclei made of protons only. Calculating the 
analogue of (\ref{rab}) for these hypothetical nuclei we define
$R^{sh}_{AB}$, which is free of isospin effects and contains shadowing 
only (pure shadowing). The ratio $R^{sh}_{AB}$ is shown at $y=0$
in Fig.~13 by the dotted lines for (a) $dAu$ and (b) $AuAu$ collisions.

It can be seen from Fig.~13 that in the central rapidity region, when 
$Q_T> 3$ GeV, $R_{AB}^{sh}$ is larger than 1 in both $AuAu$ and $dAu$ 
collisions. This means that anti-shadowing in the PDF start to be 
important at $Q_T\sim 3$ GeV. The result is consistent with what was
learned about the dominant region of $x$ earlier. Recall from Fig.~9(a)  
that the dominant region of $x$ for partons from both beams is 
about $[10^{-2},10^{-1}]$ for $Q_T=3$~GeV and $y=0$. An inspection of the EKS
nuclear PDFs\cite{Eskola:1998df} informs that $x\sim 10^{-1}$ is the 
region where anti-shadowing starts to take over shadowing. In $AuAu$ 
collisions, partons in both beams experience the same shadowing (anti-shadowing) 
effects  at central rapidity. In $dAu$ collisions, only the partons 
in the $Au$ nucleus have significant shadowing (anti-shadowing) effects. 
Therefore pure shadowing (anti-shadowing) effects are stronger in the $AuAu$ 
collisions than in the $dAu$ collisions as seen in Fig.~13.

Now let us also investigate the pure isospin effects on low-mass Drell-Yan 
production. Isospin effects originate in the difference between the parton 
distributions of neutrons and protons. Due to  this difference, the production 
cross section of Drell-Yan pairs in proton-neutron and neutron-neutron 
interactions differs from that in $pp$ collisions. To study isospin effects, 
let us introduce
\begin{equation}
R_{AB}^{iso}(Q_T) \equiv \left.
\frac{d\sigma(Z_A/A,Z_B/B)}{dQ_T^2 dy} \right/ 
\frac{d\sigma}{dQ_T^2 dy} \,\,\,  ,
\label{sigma_iso}
\end{equation}
where the numerator now does not include shadowing, i.e. standard CTEQ5M
PDFs are used without shadowing. Examining $R_{AB}^{iso}$
completes our program of separating the isospin and shadowing dependences
of Eq. (\ref{rab}).

\begin{figure}
\begin{minipage}[c]{7.8cm}
\centerline{\includegraphics[width=8cm,height=7cm]{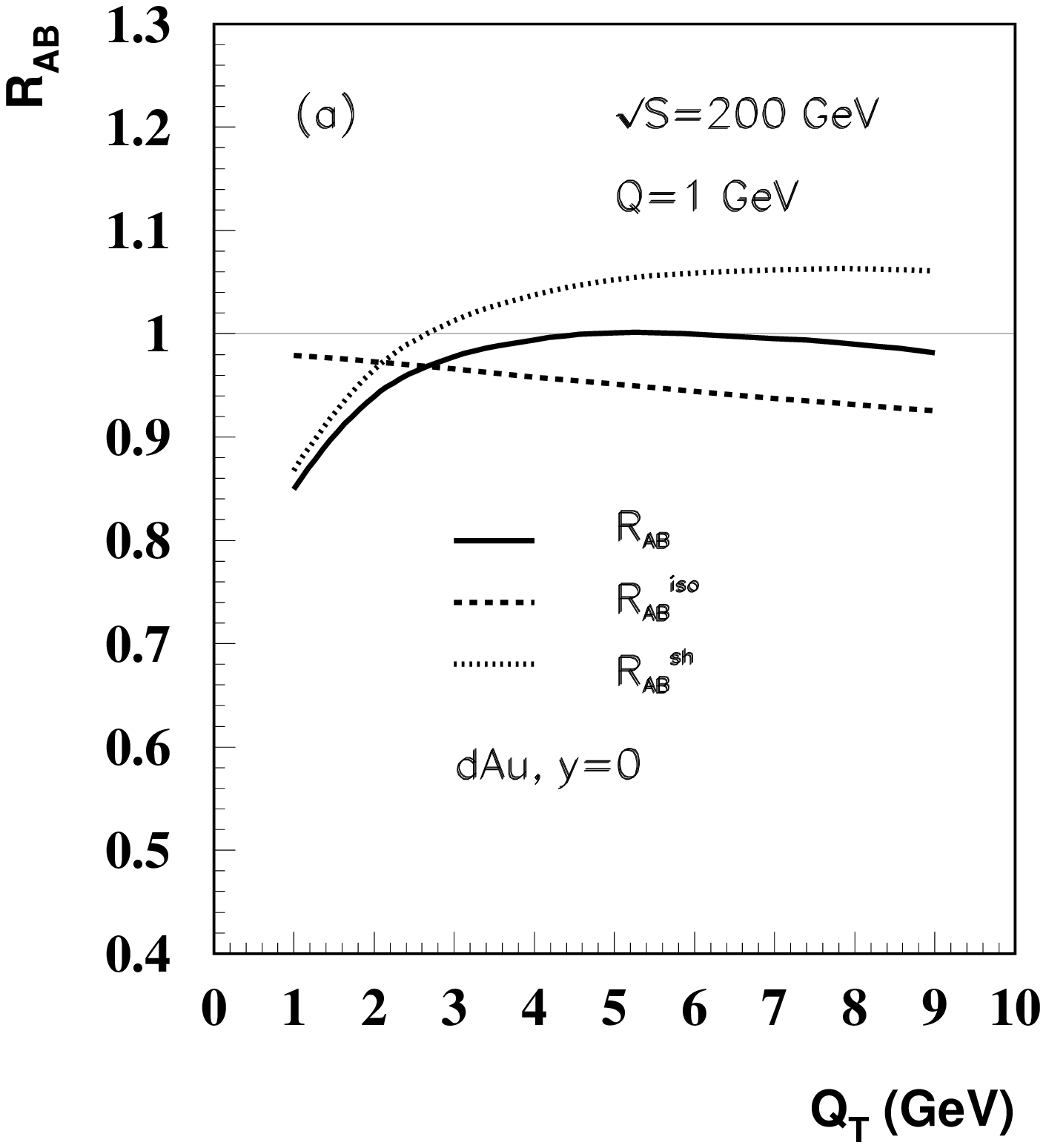}} 
\label{fig13}
\end{minipage}
\hfill
\begin{minipage}[c]{7.8cm}
\includegraphics[width=8cm,height=7cm]{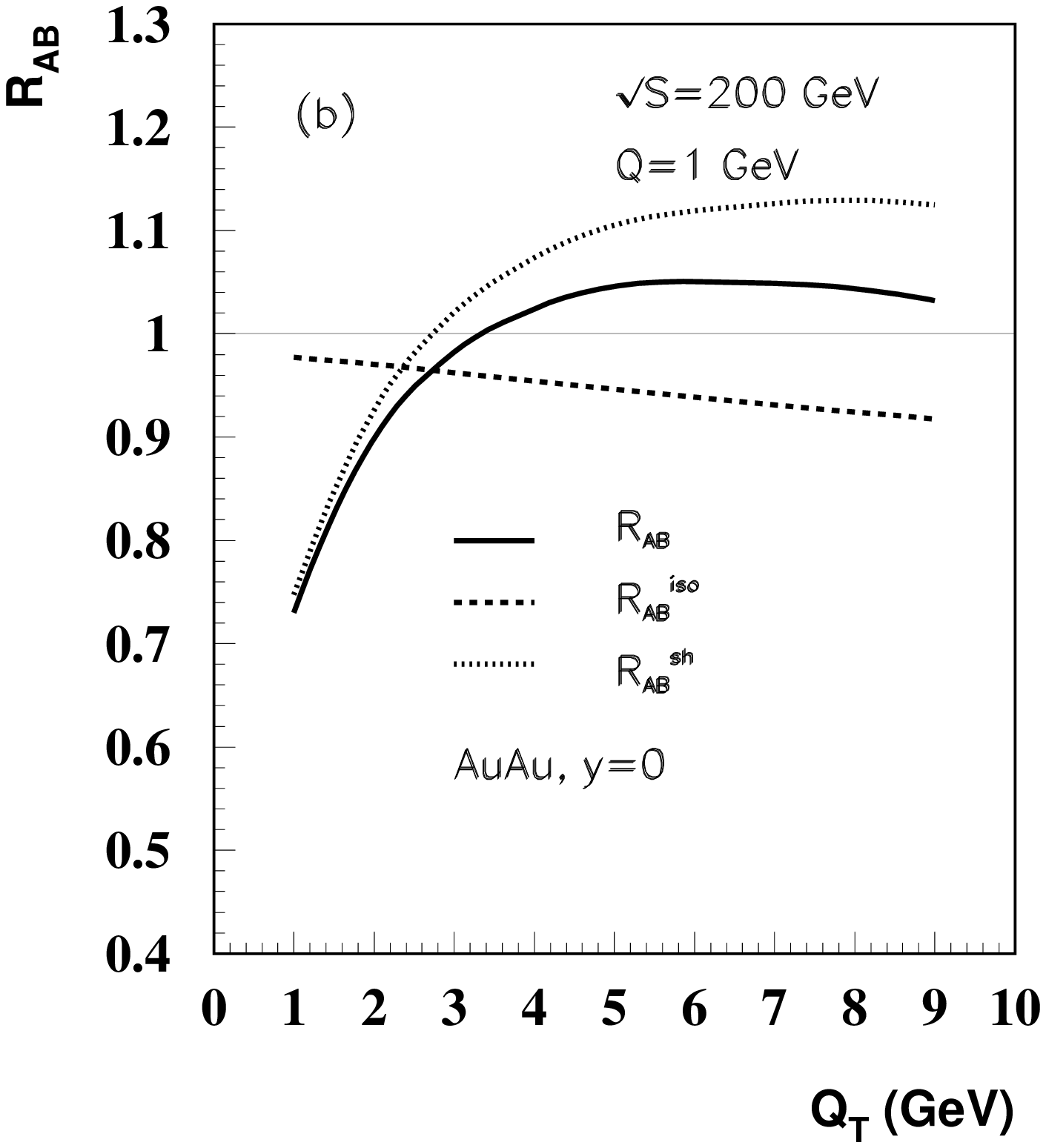} 
\end{minipage}
\vspace{0.2in}
\caption{The pure shadowing (dotted) and isospin (dashed) components of ratio $R_{AB}$
at $y=0$ and $Q=$1~GeV at RHIC for (a) $dAu$ and (b) $AuAu$ collisions. See text for 
more explanation.}
\end{figure}

Figure~13 also displays $R_{AB}^{iso}$ (dashed line) as a function of $Q_T$
in (a) $dAu$ and (b) $AuAu$ collisions at $y=0$. We see that  
$R_{AB}^{iso}<1$ in the whole transverse momentum
region $[1,10]$~GeV; isospin effects suppress the production of 
high transverse-momentum low-mass Drell-Yan pairs at RHIC. The
suppression is getting slightly stronger with increasing $Q_T$.
Since the Compton type (qg) processes are the dominant sub-processes in 
low-mass Drell-Yan production at high transverse momentum, we next consider
the isospin effects in these sub-processes to help us understand the 
suppression result shown in Fig.~13. The contribution of different 
flavor quarks is proportional to the square of their fractional charges 
(charge factors): 4/9 for $u$ quarks and 1/9 for $d$ quarks. 
Let us first look at proton-proton ($pp$) and 
proton-neutron ($pn$) collisions. In the calculation of the $pp$ Drell-Yan 
production cross section, the PDF part for the Compton-type 
processes, after including the charge factors, is proportional to 
$\Phi_g(4\Phi_u/9+\Phi_d/9 + 4\Phi_{\bar u}/9+\Phi_{\bar d}/9 + \cdots )$, 
plus an ``exchange'' term, where the gluon comes from the other proton. For $pn$,
when the gluon comes from the proton,  
we have $\Phi_g(\Phi_u/9+4\Phi_d/9 +\Phi_{\bar u}/9+4\Phi_{\bar d}/9 + \cdots)$ 
(where $\cdots$ represents the contributions which are the same for 
$pp$ and $pn$), plus an exchange term. 
In Fig.~14, we plot $\Phi_d(x,\mu)/\Phi_u(x,\mu)$ (solid),
$\Phi_{\bar d}(x,\mu)/\Phi_{\bar u}(x,\mu)$ (dashed),
and the ratio of the above two charge-weighted expressions
(dotted) as functions of $x$ at the scale $\mu^2=10$~GeV$^2$.   
For not-too-small $x$, in particular in the region of $[10^{-2}, 10^{-1}]$, 
which is the dominant region of parton $x$, valence quarks are more important 
than sea quarks. Since $\Phi_d(x,\mu)/\Phi_u(x,\mu)$ (solid) is 
itself smaller than one, the production cross section in $pn$ is smaller than 
in $pp$ collisions. The dotted line in Fig.~14 illustrates the effect.
Using a similar argument, one finds that the isospin effect in $nn$ 
collisions is a little stronger than in $pn$  collisions.  
Returning to nuclear reactions, where a large fraction of the  nucleon-nucleon 
collisions is not $pp$, but $pn$ and $nn$, we conclude that due to the charge factor 
and the fact that $\Phi_d<\Phi_u$ in the $x$ region of interest, $R^{iso}_{AB}$ 
is smaller than 1. When the transverse momentum increases, $x$ in the PDFs increases. 
Since the ratio of $\Phi_d/\Phi_u$ becomes smaller at larger $x$, $R^{iso}_{AB}$ 
decreases as $Q_T$ increases. Since the change of the proton/neutron 
ratio from $d$ to $Au$ is small for these purposes,  
$R^{iso}_{AB}$ is very similar in $dAu$ to that of $AuAu$, as Fig.~13 demonstrates.

\begin{figure}
\centerline{\includegraphics[width=10cm,height=8cm]{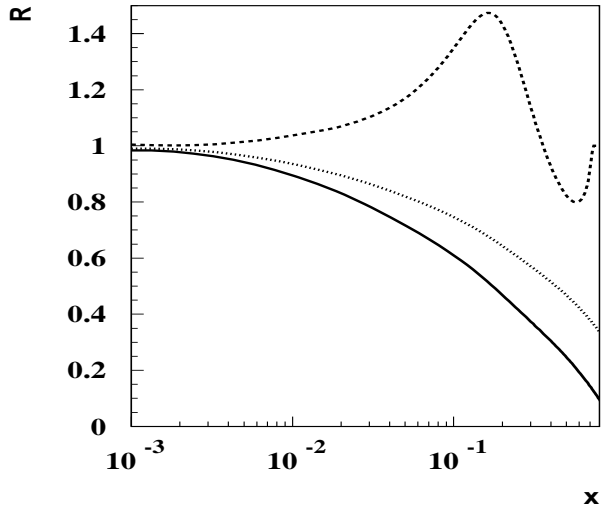}} 
\label{fig14}
\vspace{0.2in}
\caption{The ratio of the d quark and u quark PDF-s and the charge-weighted ratio
at scale $Q^2=$10~GeV$^2$. See text for more explanation.} 
\end{figure}

To summarize the above discussion,  at central rapidity $y=0$,
when $Q_T> 3$~GeV, low-mass Drell-Yan 
pair production at RHIC suffers anti-shadowing. The anti-shadowing effects 
are stronger in $AuAu$ collisions than in $dAu$ collisions; 
isospin effects suppress the production of low-mass Drell-Yan pairs and
the suppression increase with $Q_T$; isospin effects are similar in $dAu$ 
and $AuAu$ collisions. The combined effects of isospin and shadowing
lead to a larger $R_{AB}$ in $AuAu$ than in $dAu$
at RHIC energy.

Figure~15 displays results for $R_{AB}$ (solid), $R^{iso}_{AB}$  (dashed),
and $R^{sh}_{AB}$  (dotted) in the forward region for (a) $dAu$ and (b) $AuAu$ 
collisions. It is clear that the steep rise of 
$R_{AB}$ at large transverse momentum in $AuAu$ collisions 
is a consequence of the effects of the modification of the 
nuclear PDF in the ``yellow'' beam (for which $y=3$
designates the forward region); correspondingly $x_1$ enters the Fermi-motion region 
of the nuclear PDF. The large suppression due to isospin effects, which
reaches $\approx 40$\% at high transverse momentum is also from the ``yellow'' 
beam side, where the larger  $x_1$ gives a stronger $u,d$
asymmetry. To demonstrate this point, we also plot $R_{pAu}$ in
Fig.~15(a)  
(dot-dashed).  
The result $R_{pAu}\sim 1$ confirms that the large isospin effect in 
$dAu$ is from the $d$ side. It is also important to point
out the big difference between $R_{pAu}$ and $R_{dAu}$, which 
illustrates that, in regard to certain observables, 
$dAu$ collisions can be very different from $pA$ collisions.

\begin{figure}
\begin{minipage}[c]{7.8cm}
\centerline{\includegraphics[width=8cm,height=7cm]{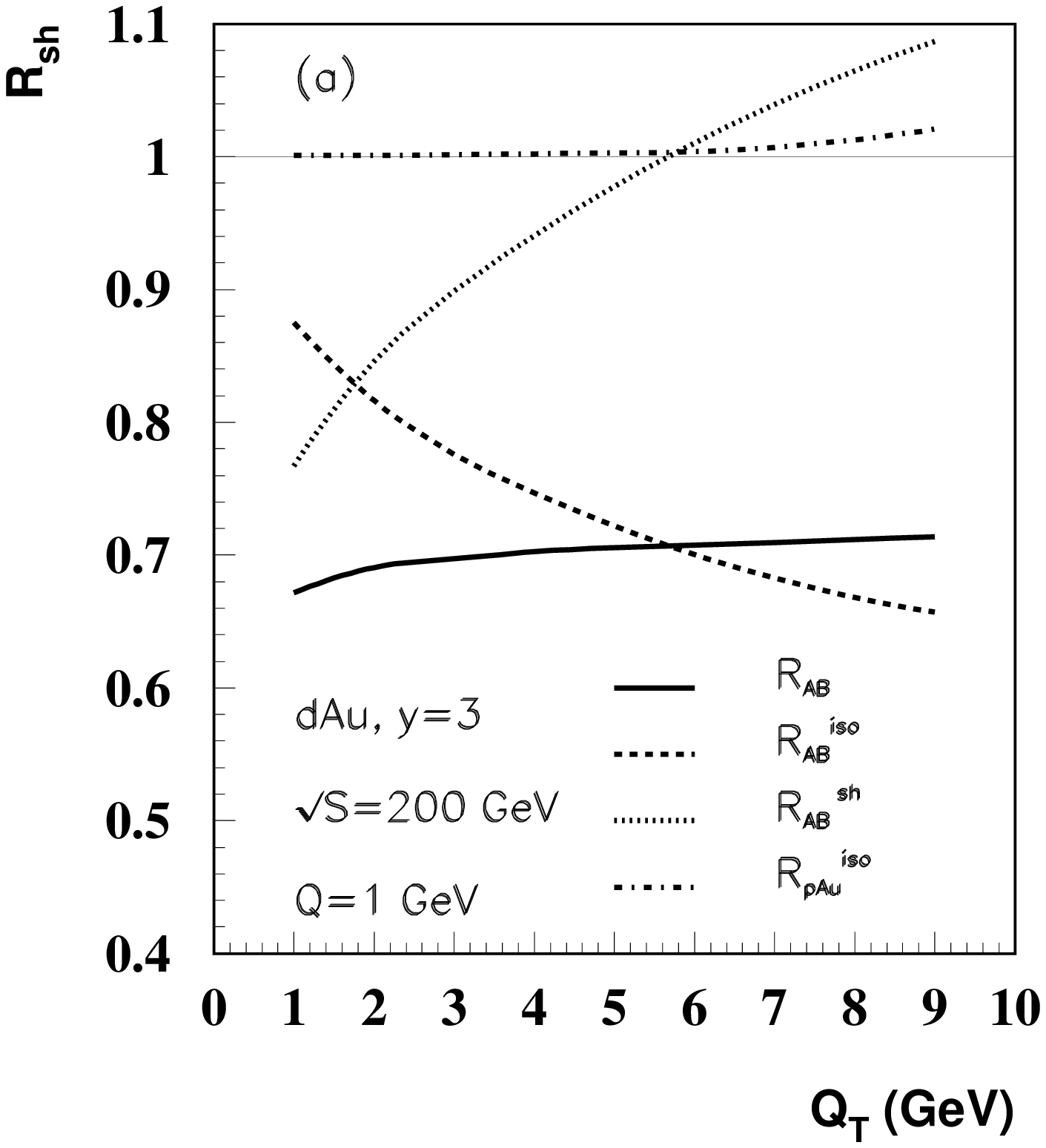}} 
\label{fig15}
\end{minipage}
\hfill
\begin{minipage}[c]{7.8cm}
\includegraphics[width=8cm,height=7cm]{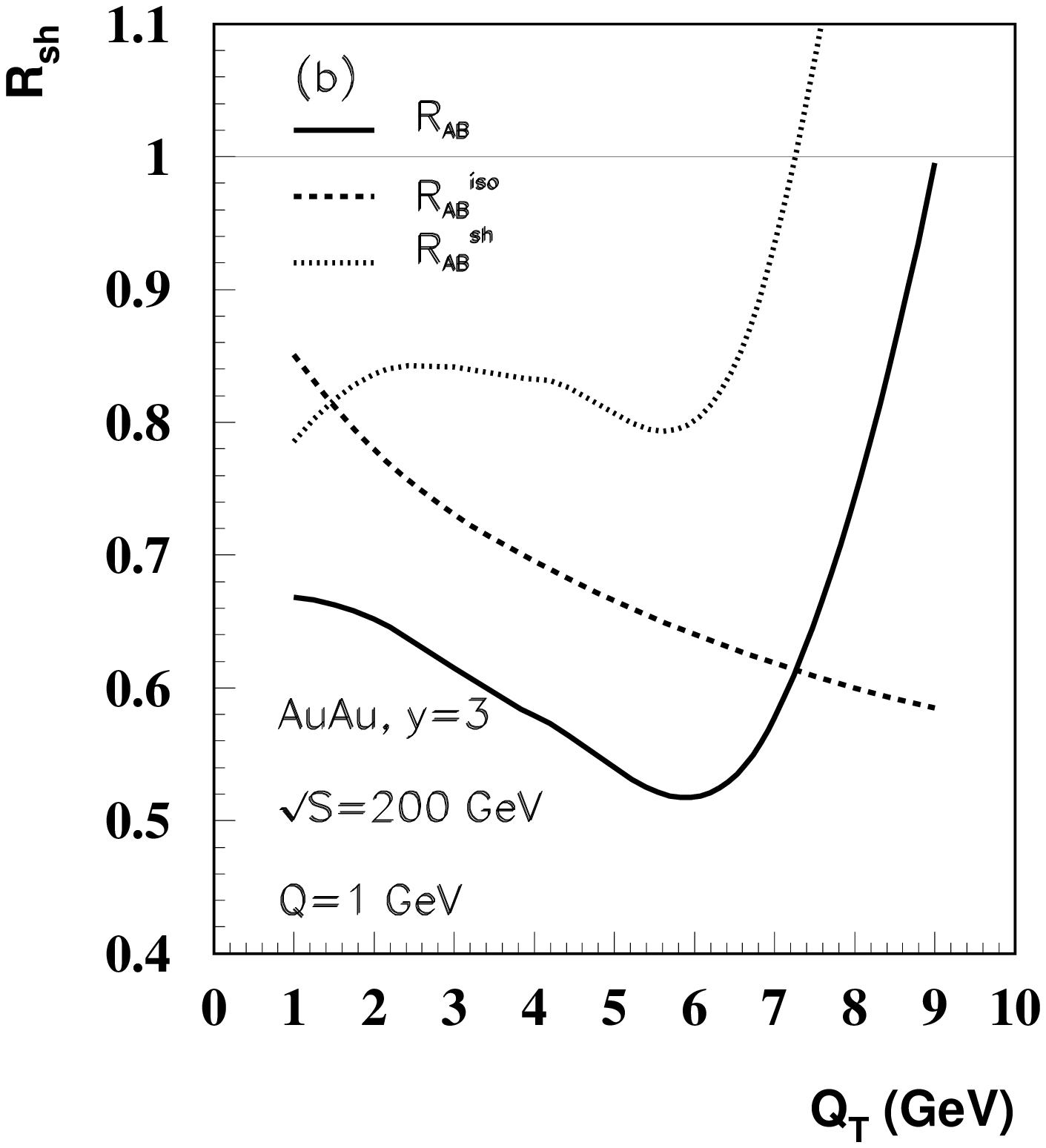} 
\end{minipage}
\vspace{0.2in}
\caption{The ratio $R_{AB}$ at RHIC in the forward region ($y=3$)
for (a) $dAu$ and (b) $AuAu$ collisions at $Q=$1~GeV as a function of $Q_T$.}
\end{figure}

Figure~16 shows $R_{AB}$ as a function of $Q_T$ at the LHC in $pPb$
and $PbPb$ collisions at both, central rapidity and forward rapidity.
In $pPb$ collisions the Drell-Yan pair production is suppressed more 
in the forward region, as expected. However, it is surprising that in 
$PbPb$ collisions, the Drell-Yan pair production in the forward region 
is actually slightly less suppressed than in the central rapidity region,
up to $Q_T \approx$ 13~GeV. This behavior is very different from what was
found at RHIC. It can be explained by the very different values of $x_1$
and $x_2$. From the central rapidity region to the forward rapidity 
region, $x_1$ increases and $x_2$ decreases. From Fig.~11, approximately,  
$x_2$ changes from on the order of $10^{-3}$ to on the order of $10^{-4}$
if one wants to produce a low-mass Drell-Yan pair not at $y=0$ but at $y=2.4$.
At the same time, $x_1$ changes from on the order of $10^{-3}$ to 
on the order of $10^{-2}$. In the EKS parameterization, the shadowing factors 
are almost constant in $[10^{-3}, 10^{-4}]$. However, shadowing effects 
decrease much in the region $[10^{-3}, 10^{-2}]$. This leads to the slight 
increase of Drell-Yan pair production in the forward region 
compared to the central rapidity region.  Gluon saturation (color glass
condensate) effects\cite{Fai:2004qu,Jalilian-Marian:2004er,Baier:2004tj}
may make the increase even larger.

\begin{figure}
\centerline{\includegraphics[width=10cm,height=8cm]{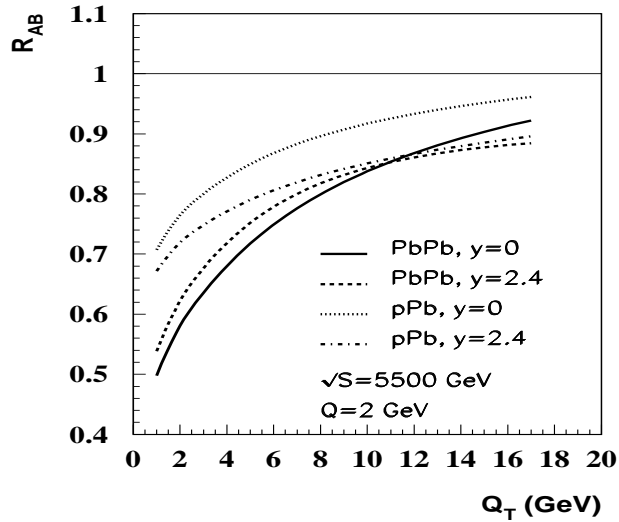}} 
\label{fig16}
\vspace{0.2in}
\caption{The ratios $R_{AB}$ at the LHC at $Q=$2~GeV for $pPb$ at $y=0$
 (dotted), $pPb$ at $y=2.4$  (dash-dot), $PbPb$ at $y=0$ (solid),
and $PbPb$ at $y=2.4$ (dashed) as a function of $Q_T$.} 
\end{figure}


\section{Summary and Conclusion}
\label{sec:concl}

In summary, low-mass Drell-Yan dilepton production 
is a potentially clean probe of small-$x$ gluons, without strong
final-state interactions.  When $Q_T>2$ GeV,   
the gluon initiated sub-processes contribute more than 
80\% to the cross section.
Unfortunately, low-mass Drell-Yan dilepton production 
suffers from low production rate at RHIC due to the Drell-Yan factor.
At the Tevatron and LHC energies, high-$Q_T$ low-mass Drell-Yan
production  is an excellent probe of the gluon distributions.

We have shown that the production rate for low-mass Drell-Yan pairs at 
Tevatron and LHC energies does not reduce for a wide range of rapidity, 
and the forward region is very sensitive to small-$x$ gluons.
Therefore, the rapidity distribution of low-mass Drell-Yan
pairs at large enough transverse momentum could be a good and clean
probe of small-$x$ gluons at collider energies.  

Because of the lack of final-state interactions and
the relatively weak initial-state nuclear-dependent power corrections
at high $Q_T$, the rapidity and transverse momentum
distribution of low-mass Drell-Yan pairs in hadron-nucleus and
nucleus-nucleus collisions should be a very good probe of the 
gluon distribution in a nucleus.  By comparing the nuclear modification
factor of $pAu$ and $dAu$, we demonstrated quantitatively that the
isospin effect is very important in extracting nuclear parton
distributions.  We have shown that low-mass Drell-Yan pairs in the forward
region at Tevatron and LHC energies provide very sensitive information
on the gluon distribution and its nuclear dependence for $x$ less than
10$^{-4}$.  With such sensitivity on gluons, low mass Drell-Yan
dilepton can be a good probe of the novel phenomena of gluon
saturation or color glass condensate in a large nucleus.
 We believe that in view of the importance of small-$x$ gluons,
the ability of low-mass Drell-Yan pairs to provide information on 
these gluon distributions as demonstrated in the present paper 
warrants the mounting of experimental efforts to measure low-mass
Drell-Yan pairs at colliders.

\vskip .5 cm

\acknowledgments
We are grateful to Mike Albrow and Mark Strikman for discussions about
small-$x$ gluon physics at CDF.
This work was supported in part by the U.S. Department of Energy,
under grants DE-FG02-86ER40251 and DE-FG02-87ER40371.  

\vspace{0.5cm}



\begin{thebibliography}{4}

\bibitem{Huston:1998jj}
J.~Huston, S.~Kuhlmann, H.~L.~Lai, F.~I.~Olness, J.~F.~Owens, D.~E.~Soper and W.~K.~Tung,
Phys.\ Rev.\ D {\bf 58}, 114034 (1998)
[arXiv:hep-ph/9801444].

\bibitem{Pumplin:2002vw}
J.~Pumplin, D.~R.~Stump, J.~Huston, H.~L.~Lai, P.~Nadolsky and W.~K.~Tung,
JHEP {\bf 0207}, 012 (2002)
[arXiv:hep-ph/0201195].

\bibitem{Martin:2003sk}
A.~D.~Martin, R.~G.~Roberts, W.~J.~Stirling and R.~S.~Thorne,
Eur.\ Phys.\ J.\ C {\bf 35}, 325 (2004)
[arXiv:hep-ph/0308087].

\bibitem{Eskola:2002yc}
K.~J.~Eskola, H.~Honkanen, V.~J.~Kolhinen, J.~W.~Qiu and C.~A.~Salgado,
Nucl.\ Phys.\ B {\bf 660}, 211 (2003)
[arXiv:hep-ph/0211239].

\bibitem{Qiu:1986wh}
J.~W.~Qiu,
Nucl.\ Phys.\ B {\bf 291}, 746 (1987).

\bibitem{Frankfurt:xz}
L.~L.~Frankfurt, M.~I.~Strikman and S.~Liuti,
Phys.\ Rev.\ Lett.\  {\bf 65} (1990) 1725.

\bibitem{Eskola:1992zb}
K.~J.~Eskola,
Nucl.\ Phys.\ B {\bf 400} (1993) 240.

\bibitem{Eskola:1998df}
K.~J.~Eskola, V.~J.~Kolhinen and C.~A.~Salgado,
Eur.\ Phys.\ J.\ C {\bf 9} (1999) 61
[arXiv:hep-ph/9807297].

\bibitem{Eskola:1998iy}
K.~J.~Eskola, V.~J.~Kolhinen and P.~V.~Ruuskanen,
Nucl.\ Phys.\ B {\bf 535} (1998) 351
[arXiv:hep-ph/9802350].

\bibitem{Hirai:2001np}
M.~Hirai, S.~Kumano and M.~Miyama,
Phys.\ Rev.\ D {\bf 64} (2001) 034003
[arXiv:hep-ph/0103208].

\bibitem{Huang:1997ii}
Z.~Huang, H.~J.~Lu and I.~Sarcevic,
Nucl.\ Phys.\ A {\bf 637} (1998) 79
[arXiv:hep-ph/9705250].

\bibitem{Li:2001xa}
S.~y.~Li and X.~N.~Wang,
Phys.\ Lett.\ B {\bf 527} (2002) 85
[arXiv:nucl-th/0110075].

\bibitem{Frankfurt:2002kd}
L.~Frankfurt, V.~Guzey, M.~McDermott and M.~Strikman,
JHEP {\bf 0202} (2002) 027
[arXiv:hep-ph/0201230].

\bibitem{Eskola:2002us}
K.~J.~Eskola, H.~Honkanen, V.~J.~Kolhinen and C.~A.~Salgado,
Phys.\ Lett.\ B {\bf 532} (2002) 222
[arXiv:hep-ph/0201256].

\bibitem{Armesto:2002ny}
N.~Armesto,
Eur.\ Phys.\ J.\ C {\bf 26} (2002) 35
[arXiv:hep-ph/0206017].

\bibitem{Accardi:2003be}
A.~Accardi {\it et al.},
arXiv:hep-ph/0308248.

\bibitem{Berger:1998ev}
E.~L.~Berger, L.~E.~Gordon and M.~Klasen,
Phys.\ Rev.\ D {\bf 58}, 074012 (1998)
[arXiv:hep-ph/9803387].

\bibitem{Berger:1990es}
E.~L.~Berger and J.~W.~Qiu,
Phys.\ Lett.\ B {\bf 248}, 371 (1990).

\bibitem{Berger:2001wr}
E.~L.~Berger, J.~W.~Qiu and X.~f.~Zhang,
Phys.\ Rev.\ D {\bf 65}, 034006 (2002)
[arXiv:hep-ph/0107309].

\bibitem{Fai:2004qu}
G.~Fai, J.~W.~Qiu and X.~f.~Zhang,
J.\ Phys.\ G {\bf 30}, S1037 (2004)
[arXiv:hep-ph/0403126].

\bibitem{Qiu:2001nr}
J.~W.~Qiu and X.~f.~Zhang,
Phys.\ Rev.\ D {\bf 64}, 074007 (2001)
[arXiv:hep-ph/0101004].

\bibitem{UA1-Vph}
C. Albajar {\it et al.}, UA1 Collaboration, Phys. Lett. {\bf B209},
397 (1988).

\bibitem{Guzey:2004zp}
V.~Guzey, M.~Strikman and W.~Vogelsang,
arXiv:hep-ph/0407201.

\bibitem{Qiu:2001zj}
J.~W.~Qiu and X.~f. ~Zhang,
Phys.\ Lett.\ B {\bf 525}, 265 (2002)
[arXiv:hep-ph/0109210].

\bibitem{Jalilian-Marian:2004er}
J.~Jalilian-Marian,
Nucl.\ Phys.\ A {\bf 739}, 319 (2004)
[arXiv:nucl-th/0402014].

\bibitem{Baier:2004tj}
R.~Baier, A.~H.~Mueller and D.~Schiff,
Nucl.\ Phys.\ A {\bf 741}, 358 (2004)
[arXiv:hep-ph/0403201].

\end{thebibliography}
\end{document}